\documentclass[12pt,preprint]{aastex}
\usepackage{epsf}
\usepackage{times}
\newcommand{\etal}{{et al}\/.}
\newcommand{\hh}{^{\rm h}}
\newcommand{\mm}{^{\rm m}}
\begin{document}
\slugcomment{Draft of \today}
\shorttitle{Magnetic fields in three FRII sources}
\shortauthors{M.J.\ Hardcastle \etal}
\title{Magnetic field strengths in the 
hotspots and lobes of three powerful FRII radio sources}
\author{M.J.\ Hardcastle, M.\ Birkinshaw}
\affil{Department of Physics, University of Bristol, Tyndall Avenue,
Bristol BS8 1TL, UK}
\author{R.A.\ Cameron, D.E.\ Harris}
\affil{Harvard-Smithsonian Center for Astrophysics, 60 Garden Street, Cambridge, MA~02138, USA}
\author{L.W.\ Looney}
\affil{Max-Planck-Institut f\"{u}r Extraterrestrische Physik (MPE), Garching, Germany}
\and
\author{D.M.\ Worrall}
\affil{Department of Physics, University of Bristol, Tyndall Avenue,
Bristol BS8 1TL, UK}
\begin{abstract}
We have made deep {\it Chandra} observations of three powerful FRII
radio sources: two quasars (3C\,263 and 3C\,351) and one radio
galaxy (3C\,330). X-ray emission from hotspots and lobes, as well as
from the active nucleus, is detected
in each source.

We model the hotspots' synchrotron spectra using VLA, BIMA and {\it
Hubble Space Telescope} data. In 3C\,263 and 3C\,330, the hotspots'
X-ray emission is at a level consistent with being synchrotron
self-Compton (SSC) emission, with a hotspot magnetic field close to
the equipartition value. In the two hotspots of 3C\,351, however, an
SSC origin for the X-rays would require the magnetic field strength to
be an order of magnitude below the equipartition value in our models:
in addition, there are offsets between the radio, optical and X-ray
emission from the secondary hotspot which are hard to explain in a
simple SSC model. We discuss the emission mechanisms that may be
responsible for these observations.

On our preferred model, the X-ray emission from the radio lobes of the
three sources is due to inverse-Compton scattering of the
microwave background radiation. If this is the case, the magnetic
field strengths in the lobes are typically about a factor 2 below the
equipartition values, assuming uniform lobe electron and magnetic
field distributions.

We detect extended X-ray emission, which we attribute to a
cluster/group environment, around 3C\,263 and 3C\,330. This detection
allows us to show that the lobes are close to pressure balance with
their surroundings, as long as no non-radiating particles contribute
to the internal pressure of the lobes.

\end{abstract}
\maketitle
\section{Introduction}
\label{intro}

{\it Chandra} has now detected a large number of X-ray features
related to the jets and hotspots of extragalactic radio sources
(see \citealt{hk02} for a recent review). The jet detections
have attracted a great deal of interest, but much of the fundamental
physics behind them remains unclear. The X-ray jets commonly seen in
FRI sources \citep*{wbh01a} probe the
high-energy particle acceleration in these objects \citep*{hbw01b}, while the few FRII jets that have been
detected \citep[e.g.,][]{smlp00} may be evidence for extremely
high bulk speeds on kiloparsec scales \citep[e.g.,][]{tmsu00}.
But the details of the processes responsible for producing the X-rays
in both classes of source are still debatable, and the issue is
confused by uncertainties about the Doppler boosting factors of the
X-ray and radio-emitting material.

In hotspots the situation is more clear-cut. These structures, which
are believed to trace a strong terminal shock or shocks at the end of
the jet in powerful FRII sources, are generally thought to be unlikely
to be moving with respect to the observer at highly relativistic
speeds \citep{bblr95, s95, al00},
although moderately relativistic motions in and beyond the hotspot
(with $v \sim 0.3c$) may be required to explain some observations
\citep{dbls97}. Their radio-to-optical spectra are often
well constrained, giving limits on the amount of synchrotron emission
expected in the X-ray, and their sizes and structures are often well
known. This means that it is possible to make a simple prediction of
the X-ray flux densities expected from inverse-Compton emission. The
dominant emission process for bright, compact hotspots which show
strong spectral steepening or a synchrotron cutoff at high
radio/optical frequencies is likely to be synchrotron-self-Compton
emission (SSC), in which the radio photons from the synchrotron
process are inverse-Compton scattered into the X-ray band by the
synchrotron-emitting electrons. A prediction of the SSC flux density
can be made if the magnetic field strength in the hotspot, and
therefore the electron number density, can be estimated. Conversely, a
measured X-ray flux density which is inferred to be SSC can be used to
estimate the magnetic field strength in the hotspot.

\begin{deluxetable}{lllll}
\tablewidth{13cm}
\tablecaption{SSC and unknown high-energy FRII hotspot detections}
%\tablewidth{10cm}
\tablehead{\colhead{Mechanism} & \colhead{Source} & \colhead{Source
type} & \colhead{Instrument}&\colhead{Reference}}
\startdata
SSC near&Cygnus A&NLRG&{\it ROSAT}&1\\
equipartition&&&{\it Chandra}&2\\
&3C\,295&NLRG&{\it Chandra}&3\\
&3C\,123&NLRG&{\it Chandra}&4\\
&3C\,196\tablenotemark{a}&Q&{\it HST}&5\\
&3C\,207&Q&{\it Chandra}&6\\
&3C\,263&Q&{\it Chandra}&This paper\\
&3C\,330&NLRG&{\it Chandra}&This paper\\[2pt]
\tableline
Unclear\tablenotemark{b}&3C\,351&Q&{\it Chandra}&7, this paper\\
\tableline
Non-SSCE&Pictor A&BLRG&{\it Einstein}&8\\
&&&{\it Chandra}&9\\
&3C\,390.3&BLRG&{\it ROSAT}&10, 11\\
&3C\,303\tablenotemark{c}&Q&{\it ROSAT}&12\\
\enddata
\tablenotetext{a}{Claimed optical SSC detection.}
\tablenotetext{b}{This source will be discussed in more detail below.}
\tablenotetext{c}{X-ray source may be a background quasar.}
\tablecomments{Source type abbreviations are as follows: NLRG,
narrow-line radio galaxy; BLRG, broad-line radio galaxy; Q, quasar.
}
\tablerefs{(1) \citealt{hcp94}; (2) \citealt{wys00}; (3)
\citealt{hnpb00}; (4) \citealt{hbw01a}; (5) \citealt{h01}; (6)
\citealt{bbcs02a}; (7) \citealt{bbcp01}; (8) \citealt{rm87}; (9)
\citealt{wys01}; (10) \citealt{p97}; (11) \citealt{hll98}; (12) \citealt{hw99}}
\label{hotspots}
\end{deluxetable}

{\it ROSAT} observations of the archetypal FRII object Cygnus A
\citep*{hcp94} showed that the X-ray emission from its hotspots
was consistent with the SSC process if the magnetic field strength was
close to the equipartition or minimum-energy values \citep{b56}.
This result is often taken as evidence for equipartition between
magnetic fields and synchrotron-emitting electrons in radio sources in
general. Other {\it ROSAT} observations, however \citep*[e.g.,][]{hll98}, showed that some hotspots had X-ray emission too bright to be
produced by the SSC mechanism with equipartition field strengths,
suggesting that a different emission process was responsible. At
the time of writing, after a number of new {\it Chandra} hotspot
detections, this dichotomy remains, as shown in Table \ref{hotspots}.
Three interesting facts about this division are immediately apparent.
Firstly, the number of objects whose emission processes are not from
SSC near equipartition (hereafter `non-SSCE' objects) is still a
significant fraction of the total. Secondly, the non-SSCE objects all
display broad emission lines, which may, after all, indicate some role
for beaming in the hotspots. And, thirdly, the non-SSCE objects all
have optical synchrotron hotspots.

The major remaining questions in this area are therefore:
\begin{enumerate}
\item \label{q1} Are magnetic field
strengths close to equipartition typical in hotspots?
\item What emission process is responsible for non-SSC sources, and
is it related to their other common properties, or to relativistic
beaming?
\end{enumerate}

Question \ref{q1} can only be addressed by new observations of
hotspots, and in the first part of this paper we report on our
deep {\it Chandra} observations of three sources selected on the basis
of bright compact radio hotspot emission, describing the emission from
their hotspots, lobes, nuclei and cluster environments. In \S7 we
will return to the second question.

The remainder of the paper is organized as follows. In \S\ref{obs} we
discuss the selection of the sample and the observations we have used.
In \S 3 we discuss the general methods of analysis we have applied: in
\S\S 4, 5 and 6 we discuss the results of applying these methods to
our three target sources. Finally, in \S\S 7 and 8, we explore the general
conclusions that can be drawn from our results.

Throughout the paper we use a cosmology with $H_0 = 65$ km s$^{-1}$
Mpc$^{-1}$, $\Omega_{\rm m} = 0.3$ and $\Omega_\Lambda = 0.7$.
Spectral indices $\alpha$ are the energy indices and are defined
throughout in the sense $S \propto \nu^{-\alpha}$. J2000.0
co-ordinates are used throughout.

\section{Sample, observations and data reduction}
\label{obs}

\begin{deluxetable}{lrrrrrr}
\tablecaption{Properties of the observed sample}
\tablehead{
Source&$z$&Scale&$S_{\rm 178\ MHz}$&Largest angular&Linear size&Galactic
$N_{\rm H}$\\
&&(kpc/arcsec)&(Jy)&size (arcsec)&(kpc)&($\times 10^{20}$ cm$^{-2}$)
}
\startdata
3C\,263&0.6563&7.5&16.6&51&380&0.91\\
3C\,330&0.5490&6.9&30.3&60&410&2.94\\
3C\,351&0.371&5.5&14.9&74&410&2.03\\
\enddata
\tablecomments{Hydrogen column densities for 3C\,263 and 3C\,351 are taken from \citet{ls95}; for 3C\,330 the quoted value is interpolated from the
database of \citet{sgwb92}.}
\label{props}
\end{deluxetable}

\subsection{Sample selection}

We selected our target objects from the 3CRR sample \citep*{lrl83},
not because the selection properties of that sample were particularly
important to our scientific goals, but because the sample members at
$z \le 1$ have almost all been imaged in the radio at high resolution
and with good sensitivity. As a consequence, the properties of the
hotspots in this sample are well known. We selected three objects with
bright, compact hotspots: 3C\,263, 3C\,351 and 3C\,330. To predict an
approximate {\it Chandra} hotspot count rate, assuming equipartition,
we used a simple SSC model for each source, and then proposed
observations which would allow a high-significance detection of the
hotspots unless the magnetic field was much greater than the
equipartition value. The general properties of our target sources are
listed in Table \ref{props}.

\begin{deluxetable}{llrrr}
\tablecaption{{\it Chandra} observations}
\tablewidth{12cm}
\tablehead{Source&Date&Livetime&Effective time&Observatory\\
&&(s)&(s)&roll angle}
\startdata
3C\,263&2000 Oct 28&49190&44148&37\degr\\
3C\,351&2001 Aug 24&50920&45701&254\degr\\
3C\,330&2001 Oct 16&44083&44083&319\degr\\
\enddata
\tablecomments{
The effective time quoted is the livetime after high-background
filtering. The satellite roll angle $\theta_{\rm roll}$ determines the
position angle on the sky of any readout streak; the position angles
(defined conventionally as north through east) are $360 - \theta_{\rm
roll}$, $180 - \theta_{\rm roll}$. }
\label{xobs}
\end{deluxetable}

\subsection{X-ray observations}

The three targets were observed with {\it Chandra} as shown in Table
\ref{xobs}. The standard ACIS-S observing configuration was used for
3C\,263 and 3C\,330. For 3C\,351 we used a standard half-chip subarray
and only read out four chips, reducing the frame time to 1.74 s, in
order to reduce pileup in the bright X-ray nucleus. Subsequent
processing was carried out using {\sc ciao} 2.2 and {\sc caldb
2.9}. All three observations were processed identically. The initial
processing steps consisted of removing high-background intervals
(significant for 3C\,263 and 3C\,351)
and generating a new level 2 events file with the latest calibration
applied and with the 0.5-pixel randomization removed. After aligning
the X-ray and radio cores to correct for small offsets between the
radio and X-ray images, which we attribute to uncertainties in the
{\it Chandra} aspect determination (discussed in more detail below),
we used smoothed images made in the 0.4--7.0-keV bandpass to define
extraction regions for the various detected X-ray components.

\begin{deluxetable}{lllrrrrl}
\tablecaption{VLA observations used in the analysis}
\tabletypesize{\scriptsize}
\tablehead{Source&Program&Date&Frequency&Array&Observing time&Resolution&Comment\\
&ID&&(GHz)&&(s)&('')}
\startdata
3C\,263&AH737&2001 Jan 03&14.94&A&4390&0.14 $\times$ 0.10&Our observations\\
&AP380&2000 Apr 27&4.86&C&1500&4.67 $\times$ 3.31&Map supplied by L.M.\ Mullin\\
&AL270&1992 Oct 31&1.44&A&2930&1.51 $\times$ 1.12&Archival $uv$ data\\
&AW249&1991 Aug 31&8.26&A&1150&0.28 $\times$ 0.19&Archival $uv$ data\\
&AW249&1991 Nov 13&8.26&B&1160&0.87 $\times$ 0.69&Archival $uv$ data\\
&AB454&1987 Jul 11&4.87&A&23940&0.36&Map supplied by R.A.\ Laing (1)\\
&AB454&1987 Dec 06&4.87&B&14300&1.00&Map supplied by R.A.\ Laing (1)\\[3pt]
3C\,330&AP331&1996 Dec 15&8.47&A&5050&0.30&Map supplied by J.M.\ Riley (2)\\
&AP331&1997 Apr 13&8.47&B&2920&&(as above)\\
&AL200&1989 Sep 19&8.41&C&10170&2.50&Map supplied by J.M.\ Riley (2)\\
&AL200&1989 Nov 24&8.41&D&2570&&(as above)\\
&AM213&1987 Aug 17&1.49&A&2050&1.52 $\times$ 1.18&Archival $uv$ data\\
&AA114&1991 Jan 24&14.94&C&2850&1.49 $\times$ 1.13&Archival $uv$ data\\[3pt]
3C\,351&LAIN&1982 Mar 11&1.42&A&3190&1.85&3CRR Atlas\tablenotemark{a}\
\ (3)\\
&AL146&1987 Nov 25&1.47&B&3370&&(as above)\\
&AW249&1991 Jun 30&8.06&A&2360&0.33 $\times$ 0.23&Archival $uv$ data\\
&AW249&1991 Nov 13&8.06&B&2360&&Archival $uv$ data\\
&AW249&1991 Jan 24&8.06&C&2350&3.0&Map supplied by J.M.\ Riley (2)\\
&AP331&1998 Jan 24&8.47&D&810&&(as above)\\
&AL43&1983 May 06&14.96&C&3500&1.36 $\times$ 1.10&Archival $uv$ data\\
&LAIN&1982 May 30&14.96&A&1150&0.15 $\times$ 0.10&Archical $uv$ data\\
\enddata
\label{vla}
\tablenotetext{a}{The 3CRR Atlas can be found online at
http://www.jb.man.ac.uk/atlas/.}
\tablecomments{The resolutions
quoted are the FWHM of the restoring circular Gaussian, if a circular
beam was used, or otherwise the major and minor FWHM of the restoring
elliptical Gaussian: as maps were made from combinations of different
observations in several cases, not all individual observations have
corresponding resolutions quoted. Observing times are taken from the
VLA archive and do not include the effects of any required data
flagging.}
\tablerefs{(1) \citealt{bhlb94}; (2) \citealt{grpa02}; (3) \citealt{lp91}.}
\end{deluxetable}

\subsection{Radio data}

Radio observations are required in order to model the expected SSC
emission from the hotspot. When it became apparent that the
small-scale structure of 3C\,263's hotspot was not well enough
constrained for our purposes, the Very Large Array (VLA) schedulers
kindly allowed us to make a short observation of the source at 15 GHz
in A configuration. In addition to this new dataset, we obtained $uv$
datasets for all sources from the archives with the permission of the
original observers, while some other observers supplied us with maps
or $uv$ data directly. A complete list of the VLA observations we
used is given in Table \ref{vla}. In all cases where we reduced the
data ourselves, standard procedures were followed within {\sc aips},
using several iterations of phase self-calibration sometimes followed
by one amplitude self-calibration step. Errors in the fluxes derived
from the VLA data are dominated in all cases by the uncertainties in
primary flux calibration, nominally around 2\%.

\subsection{Millimeter-wave observations}

Observations in the millimeter band are also useful to this work, as
electrons with energies around the low-energy cutoff inferred in some
hotspots ($\gamma \sim 1000$) scatter photons at millimeter
wavelengths into the X-ray. We therefore observed all three of our
targets with the Berkeley-Illinois-Maryland Association (BIMA) array
\citep{wtpw96}. The observations were made in the B configuration
of BIMA, using two 800-MHz channels centered on 83.2 and 86.6 GHz,
which were combined to give an effective observing frequency of 84.9
GHz. Observing times and dates are given in Table \ref{bimaobs}. The
data were reduced in {\sc miriad} and final imaging was carried out
within {\sc aips}. In all cases the target hotspots were detected;
85-GHz radio cores were also detected in 3C\,263 and 3C\,351. The
resolution of BIMA in this configuration and frequency is $\sim 3''$,
so the observations allow us to distinguish between hotspot and lobe
emission and to separate double hotspots but not to comment on
structure in individual hotspots. The BIMA images will be
presented elsewhere.

\begin{deluxetable}{llll}
\tablecaption{BIMA observations}
\tablewidth{8cm}
\tablehead{Source&Date&Duration&On-source\\&&(h)&time (h)}
\startdata
3C\,263&2002 Feb 03&8.0&4.0\\
3C\,330&2002 Feb 08&4.8&2.5\\
3C\,351&2002 Feb 10&7.5&3.9\\
\enddata
\label{bimaobs}
\tablecomments{The duration given is the entire length of the run,
including slewing and calibration: this indicates the $uv$ coverage
obtained. The on-source time is the time spent observing the target,
and indicates the sensitivity of the observations.}
\end{deluxetable}

\subsection{Optical observations}

\begin{deluxetable}{llllrr}
\tablecaption{{\it HST} WFPC2 observations used in the analysis}
\tablewidth{13cm}
\tablehead{Source&ID&Date&Filter&Duration&Observatory\\
&&&&(s)&roll angle}
\startdata
3C\,263&U2SE0201&1996 Feb 18&F675W&1000&28.68\\
3C\,330&U3A14X01&1996 Jun 03&F555W&600&333.46\\
3C\,351&U2X30601&1995 Nov 30&F702W&2400&54.73\\
\enddata
\label{hst}
\tablecomments{The satellite roll angle quoted is the measured position angle
on the sky of the {\it HST}'s V3 axis, north through east. The
orientation of the science images on the sky (the position angle of
the detector $y$ axis) is the sum of this angle
and the angle between the V3 axis and the detector $y$ axis, $\sim
224.6\degr$ for these observations.}
\end{deluxetable}

For optical information, which tells us about the high-energy
synchrotron spectrum, the archival {\it Hubble Space Telescope}
({\it HST}) WFPC2 datasets listed in Table \ref{hst} were used. Cosmic
ray rejection was performed (where necessary) within {\sc iraf}, and
flux densities or upper limits were measured using standard
small-aperture photometry techniques in {\sc aips}. As with the X-ray
data, the optical data were aligned with the radio data by shifting
the optical co-ordinates to align the radio nucleus with the optical
peak. We discuss residual astrometric uncertainties in more detail
below.

\section{Analysis methods}

\subsection{X-ray spectral extraction}

\begin{figure*}
\begin{center}
\begin{minipage}{10cm}
\epsscale{0.7}
\plotone{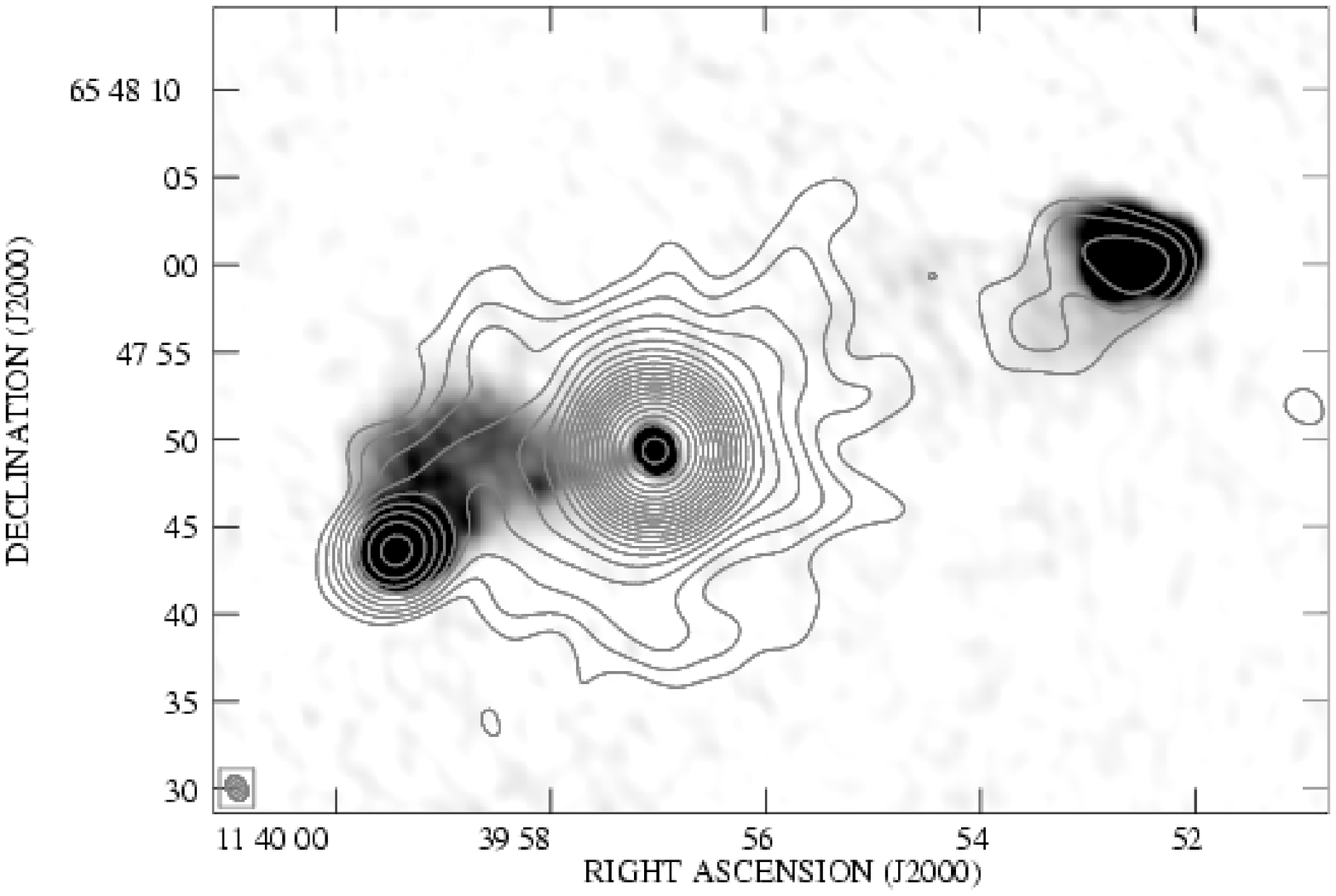}
\plotone{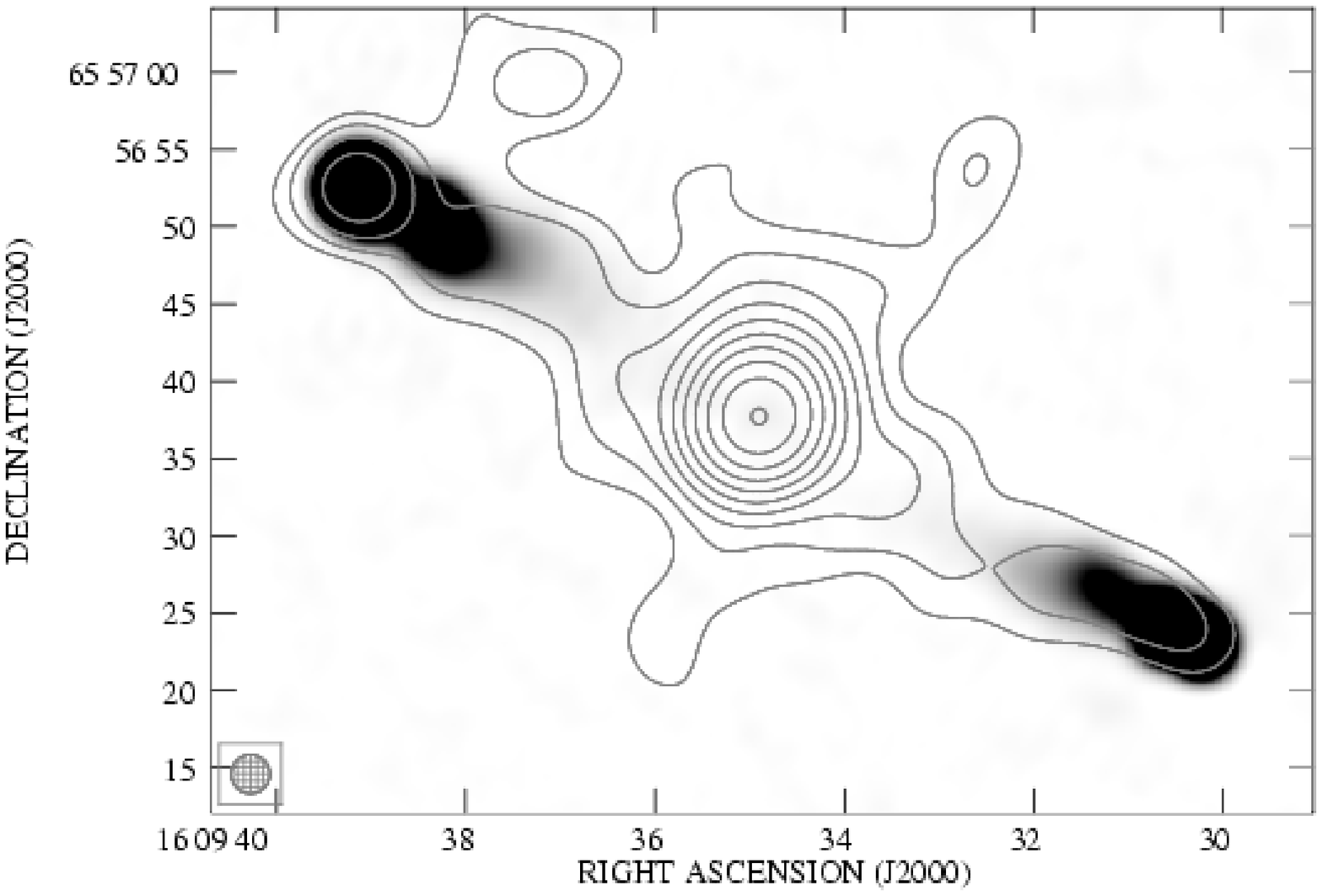}
\plotone{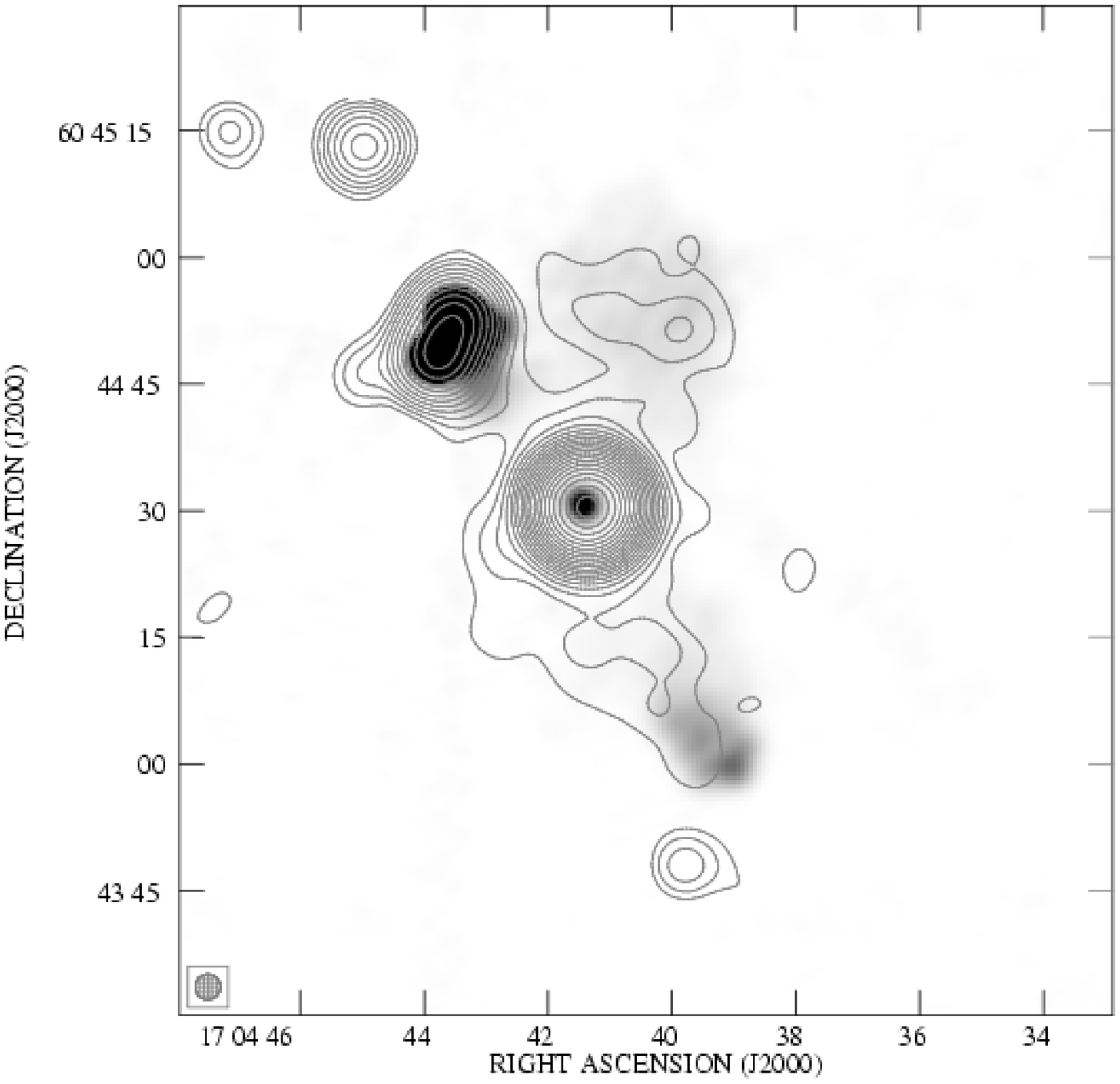}
\end{minipage}
\end{center}
\epsscale{1.0}
\caption{The three X-ray sources. From top to bottom: 3C\,263, 3C\,330
and 3C\,351. The contours show the X-ray data, smoothed with a
Gaussian of FWHM $3\farcs5$ (for 3C\,263) and $6\arcsec$ (for the other
two sources): the contour intervals are logarithmic in steps of
$\sqrt{2}$ and the lowest contour in each figure is the $3\sigma$
level as defined by \citet{h01}. The objects to the NE and S of
3C\,351 are background point sources. The greyscales show radio maps.
The map of 3C\,263 is a 1.4-GHz VLA image with $1\farcs5 \times
1\farcs1$ resolution; black is 10 mJy beam$^{-1}$. The 3C\,330 and
3C\,351 maps are 8.4-GHz VLA images, from \citet{grpa02}, with
3\arcsec\ resolution; black is 10 mJy beam$^{-1}$.}
\label{montage}
\end{figure*}

\begin{figure*}
\begin{center}
\begin{minipage}{10cm}
\epsscale{0.7}
\plotone{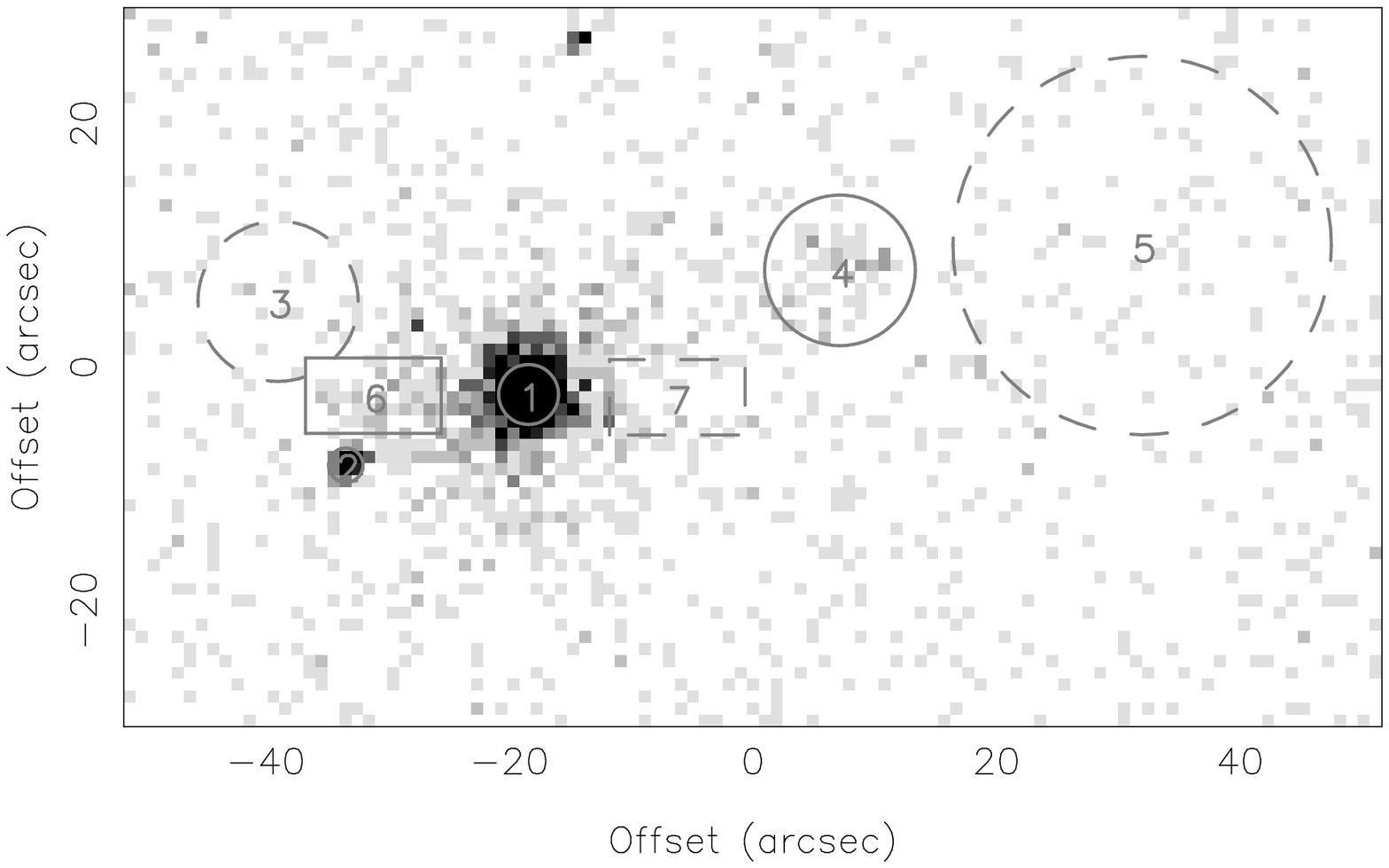}
\plotone{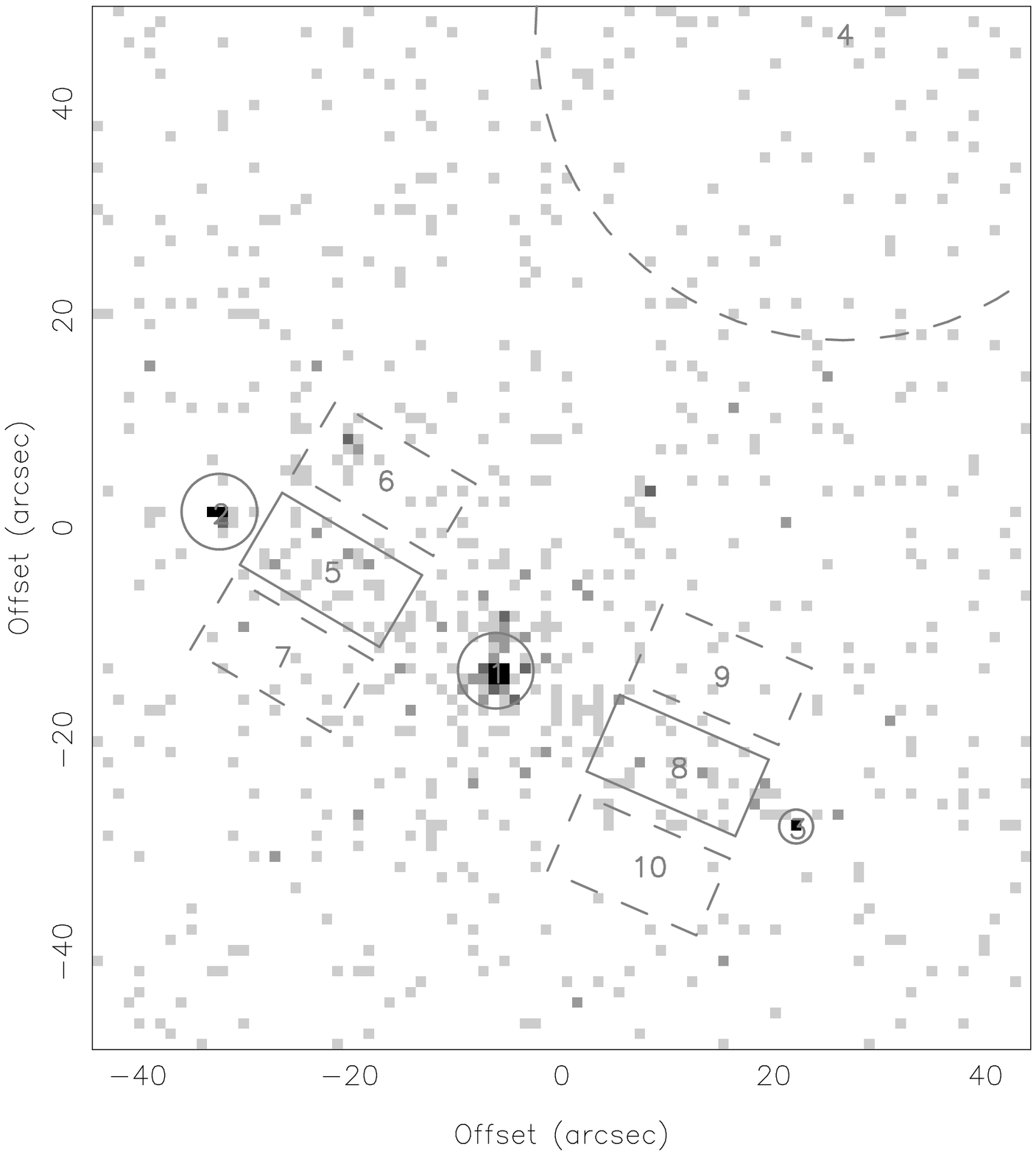}
\plotone{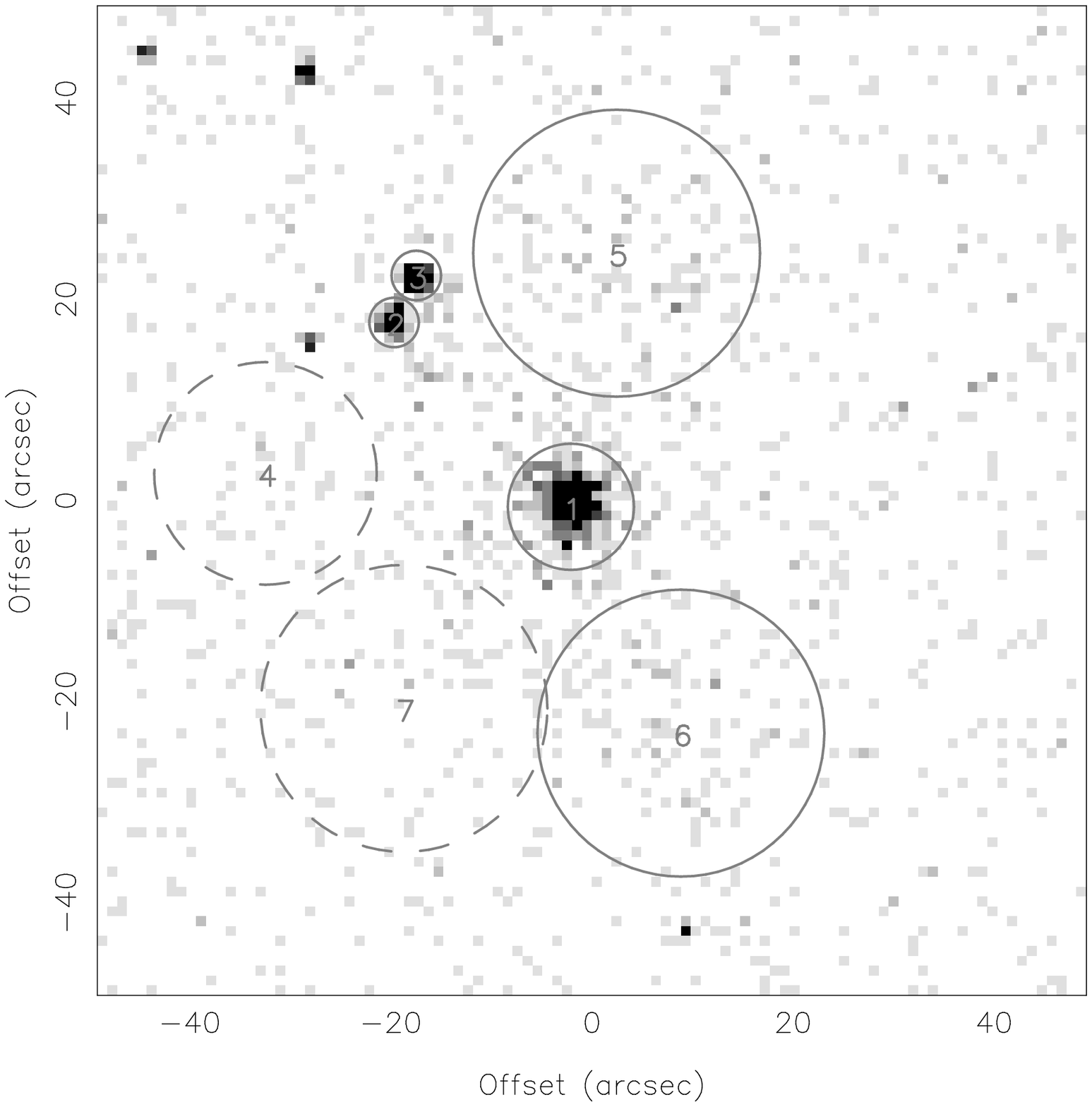}
\end{minipage}
\end{center}
\epsscale{1.0}
\caption{Extraction regions used in the three radio sources. From top
to bottom: 3C\,263, 3C\,330 and 3C\,351. Each image shows the 0.4-7
keV events binned by a factor of 2 (i.e. to 0\farcs984 pixels); black
is 8, 5 and 8 counts per bin respectively. Solid lines show source
regions, and dashed lines background regions. Numbers refer to the
regions listed in Table \ref{expos}. For clarity, a few regions have
been omitted.}
\label{regions}
\end{figure*}

We detected X-ray emission from the nucleus, from one or more
hotspots, and from the radio lobes of each source, as shown in Fig.\
\ref{montage} and discussed in the following sections. For each of
these components, spectra were extracted (from the regions listed in
Table \ref{expos}) and illustrated in Fig.\ \ref{regions} using the
{\sc ciao} script {\sc psextract}\footnote{This script extracts
spectra from regions of the events file and a chosen background, and
generates appropriate response files. See
http://cxc.harvard.edu/ciao/ahelp/psextract.html.} and analysed with
{\sc xspec}. Particular care was taken in the case of the lobe
extractions to ensure that the background regions used were at similar
or identical distances from the nucleus (Fig.\ \ref{regions}); this
minimizes systematic effects due to the radially symmetrical PSF or
any radially symmetrical cluster emission. In all cases, spectra were
binned such that each bin contained $\ga 20$ net counts. Fits were
carried out in the 0.4--7.0 keV energy band. The models used for each
source are discussed in detail in the following sections; the results
of fitting these models are tabulated for each source in Table
\ref{restab}. Errors quoted in the Table or in the text correspond to
the $1\sigma$ uncertainty for one interesting parameter, unless
otherwise stated.


\begin{deluxetable}{lllll}
\tablecaption{Spatial regions used in extraction}
\tabletypesize{\scriptsize}
\tablewidth{14cm}
\tablehead{Source&Component&Central position&Geometry&Number}
\startdata
3C\,263&Core&$11\hh39\mm
57\fs025$, $+65\degr47'49\farcs36$&Circle, $r = 5$ pixels&1\\
&SE hotspot (IJK)&$11\hh39\mm59\fs457$, $+65\degr47'43\farcs59$&Circle, $r = 2.9$ pixels&2\\
&SE hotspot (K)&$11\hh39\mm59\fs492$, $+65\degr47'43\farcs47$&Circle, $r=1.4$ pixels\\
&SE hotspot background&$11\hh40\mm00\fs361$, $+65\degr47'57\farcs06$&Circle, $r=13.4$ pixels&3\\
&NW hotspot (limit region)&$11\hh39\mm52\fs857$, $+65\degr47'59\farcs57$&Circle,
$r=12.6$ pixels&4\\
&NW lobe&$11\hh39\mm52\fs857$, $+65\degr47'59\farcs57$&Circle, $r=12.6$ pixels&4\\
&NW lobe background&$11\hh39\mm48\fs821$, $+65\degr48'01\farcs59$&Circle, $r=31.6$ pixels&5\\
&Jet (limit region)&$11\hh39\mm59\fs090$, $+65\degr47'49\farcs30$&Box, $22.6 \times 12.6$
pixels, $90\degr$&6\\
&SE lobe&$11\hh39\mm59\fs090$, $+65\degr47'49\farcs30$&Box, $22.6 \times 12.6$
pixels, $90\degr$&6\\
&SE lobe background&$11\hh39\mm55\fs030$, $+65\degr47'49\farcs16$&Box, $22.6 \times 12.6$
pixels, $90\degr$&7\\[4pt]
3C\,330&Core&$16\hh09\mm34\fs896 +65\degr56'37\farcs69$&Circle, $r = 7.3$ pixels&1\\
&NE hotspot&$16\hh09\mm39\fs155$, $+65\degr56'52.79$&Circle, $r=7.3$ pixels&2\\
&SW hotspot&$16\hh09\mm30\fs255$, $+65\degr56'23.07$&Circle, $r= 3.25$ pixels&3\\
&Hotspot/core background&$16\hh09\mm41\fs548$, $+65\degr56'13.74$&Circle, $r =
65.5$ pixels&4\\
&NE lobe&$16\hh09\mm37\fs433$, $+65\degr56'47.29$&Box, $31.1\times 16.0$ pixels, $59.5\degr$&5\\
&NE lobe background&$16\hh09\mm36\fs608$, $+65\degr56'55.90$&Box, $31.1 \times
16.0$ pixels, $59.5\degr$&6\\
&&$16\hh09\mm38\fs198$, $+65\degr56'39.30$&Box, $31.1 \times 16.0$ pixels, $59.5\degr$&7\\
&SW lobe&$16\hh09\mm32\fs083$, $+65\degr56'28.83$&Box, $31.1 \times 16.0$
pixels, $66.5\degr$&8\\
&SW lobe background&$16\hh09\mm31\fs419$, $+65\degr56'37.44$&Box, $31.1 \times 16.0$
pixels, $66.5\degr$&9\\
&&$16\hh09\mm32\fs687$, $+65\degr56'19.49$&Box, $31.1 \times 16.0$
pixels, $66.5\degr$&10\\[4pt]
3C\,351&Core&$17\hh04\mm41\fs364$, $+60\degr44'30\farcs46$&Circle,
$r=12.8$ pixels&1\\
&Core background&$17\hh04\mm41\fs364$, $+60\degr44'30\farcs46$&Annulus,
$r=12.8..25$ pixels\\
&N hotspot (J)&$17\hh04\mm43\fs781$, $+60\degr44'48\farcs20$&Circle, $r=5$ pixels&2\\
&N hotspot (L)&$17\hh04\mm43\fs479$, $+60\degr44'52\farcs88$&Circle, $r=5$ pixels&3\\
&N hotspots background&$17\hh04\mm45\fs525$, $+60\degr44'33\farcs19$&Circle, $r=22.5$ pixels&4\\
&S hotspot (limit region)&$17\hh04\mm39\fs754$, $+60\degr44'04\farcs91$&Circle, $r=16$ pixels\\
&N lobe&$17\hh04\mm40\fs760$, $+60\degr44'55\farcs09$&Circle, $r=29$ pixels&5\\
&S lobe&$17\hh04\mm39\fs888$, $+60\degr44'07\farcs37$&Circle, $r=29$ pixels&6\\
&Lobe background&$17\hh04\mm43\fs646$, $+60\degr44'09\farcs83$&Circle, $r=29$ pixels&7\\
\enddata
\tablecomments{
Positions quoted are corrected to the frame defined by the radio
observations, as discussed in the text. One {\it Chandra} pixel is
$0\farcs492$. `Limit regions' are the regions used to determine a
background count rate for an upper limit on a compact component;
smaller detection regions are then used to derive the upper limit, as
discussed in the text. Box
regions are given as long axis $\times$ short axis, position angle of
long axis (defined north through east). Numbers refer to the regions
shown in Fig.\ \ref{regions}.}
\label{expos}
\end{deluxetable}

\begin{deluxetable}{llrrp{2.9cm}rrr}
\tablecaption{X-ray properties of source components}
\tabletypesize{\scriptsize}
\tablehead{
Source&Component&Net count rate&Net counts&Spectral fit&$\chi^2/n$&2--10
keV flux&1-keV flux\\
&&(0.4--7.0 keV, &&&&($\times 10^{-15}$ ergs&density (nJy)\\
&&s$^{-1}$)&&&&cm$^{-2}$ s$^{-1}$)
}
\startdata
3C\,263&Core\tablenotemark{a}&$0.163 \pm 0.002$&$7218 \pm 85$&$\alpha = 0.75 \pm
0.06$&229/199&$1700 \pm 300$&$290 \pm 30$\\
       &SE hotspot (IJK)&$(2.5 \pm 0.5) \times 10^{-3}$&$110 \pm 22$&$\alpha=1.2 \pm
0.2$&1.4/3&$5.1 \pm 0.5$&$1.7 \pm 0.2$\\
&SE hotspot (K)&$(1.6 \pm 0.2) \times 10^{-3}$&$69 \pm 8$&$\alpha = 1.0 \pm
0.3$&0.64/1&$3.6 \pm 0.4$&$1.0 \pm 0.1$\\
&NW hotspot (B)\tablenotemark{b}&$< 1 \times 10^{-4}$&$<4$&...&...&...&$\la 0.06$\\
&Jet\tablenotemark{b}&$< 5 \times 10^{-4}$&$<20$&...&...&...&$\la 0.3$\\
&NW lobe&$(1.2 \pm 0.2) \times 10^{-3}$&$53 \pm 7$&$\alpha = 0.4 \pm 0.2$&0.98/1&$7.8 \pm 1.8$&$0.8 \pm 0.2$\\
&SE lobe\tablenotemark{c}&$(7.7 \pm 2.2) \times 10^{-4}$&$34 \pm 10$&...&...&...&$0.5 \pm 0.2$\\[4pt]
3C\,330&Core\tablenotemark{d}&$(3.4 \pm 0.5) \times 10^{-3}$&$150 \pm 22$&$\alpha_1 = 0.8_{-0.4}^{+0.5} + [N_{\rm H} =
(4\pm2)\times 10^{23}$ cm$^{-2}$, $\alpha_2 =
1.6^{+2.7}_{-1.3}]$&1.8/2&$47 \pm 7$&$1.5 \pm 0.2$\\
&NE hotspot\tablenotemark{e}&$(5.0 \pm 1.1) \times 10^{-4}$&$22 \pm 5$&...&...&...&$0.34 \pm 0.07$\\
&SW hotspot\tablenotemark{e}&$(1.3 \pm 0.6) \times 10^{-4}$&$6 \pm 3$&...&...&...&$0.09 \pm 0.04$\\
&NE lobe\tablenotemark{e}&$(4.1 \pm 1.5)\times 10^{-4}$&$18 \pm 7$&...&...&...&$0.28 \pm 0.10$\\
&SW lobe\tablenotemark{e}&$(4.7 \pm 1.2) \times 10^{-4}$&$21 \pm 5$&...&...&...&$0.32 \pm 0.08$\\[4pt]
3C\,351&Core\tablenotemark{f}&$0.160 \pm 0.002$&$7310 \pm
91$&$\alpha_1 = 2.5 \pm 0.3 + [N_{\rm H}
= (1.1 \pm 0.1) \times 10^{22}$ cm$^{-2}$, $\alpha_2 = 0.46 \pm
0.05]$&304/212&$1560 \pm 20$&$14 \pm 3$\\
&N hotspot (J)&$(6.5 \pm 0.5) \times 10^{-3}$&$297 \pm 23$&$\alpha = 0.5 \pm
0.1$&8/11&$36 \pm 4$&$4.3 \pm 0.3$\\
&N hotspot (L)&$(5.2 \pm 0.5) \times 10^{-3}$&$238 \pm 23$&$\alpha = 0.85 \pm
0.1$&17/9&$17 \pm 2$&$3.4 \pm 0.4$\\
&S hotspot\tablenotemark{g}&$<0.7 \times 10^{-4}$&$<3$&...&...&...&$\la 0.05$\\
&N lobe\tablenotemark{h}&$(1.3 \pm 0.3) \times 10^{-3}$&$59 \pm 14$&$\alpha = 0.6 \pm
0.8$&0.6/1&$8 \pm 2$&$1.1 \pm 0.3$\\
&S lobe\tablenotemark{h}&$(0.9 \pm 0.3) \times 10^{-3}$&$41 \pm 14$&$\alpha = 0.6 \pm
0.8$&0.6/1&$5 \pm 2$&$0.7 \pm 0.3$\\
\enddata
\tablenotetext{a}{Source
affected by pileup, see text; values tabulated are derived from the
{\sc xspec} pileup model, and errors quoted are $1\sigma$ for two
interesting parameters.}
\tablenotetext{b}{Flux density estimated from count rate on
the basis of the SE hotspot's spectrum; see text.}
\tablenotetext{c}{Flux density
estimated from count rate on the basis of the NW lobe's spectrum.}
\tablenotetext{d}{Flux density quoted for low-energy power law ($\alpha_1$) only.}
\tablenotetext{e}{Flux density
estimated assuming a power law with $\alpha = 0.5$ and Galactic
absorption.}
\tablenotetext{f}{Flux density quoted for unabsorbed power law only. See the text for discussion of other models fitted to
these data.}
\tablenotetext{g}{Flux density estimated from the spectrum of hotspot
J.}
\tablenotetext{h}{Spectral fits combine N and S lobes.}
\tablecomments{2-10 keV fluxes quoted are absorbed, observer-frame
values. The observer-frame 1-keV flux densities tabulated are corrected for
Galactic absorption, which is fixed to the previously measured value
(Table \ref{props}) in all fits. Errors quoted are $1\sigma$ for one
interesting parameter.}
\label{restab}
\end{deluxetable}

\subsection{Intra-cluster medium emission}

Our deep {\it Chandra} observations also allow us to search for
extended, cluster-related X-ray emission. Extended emission, on scales
larger than the radio lobes, was visible by eye in the observations of
3C\,263 and 3C\,330. To characterize this extended emission we masked
out the readout streak and the non-nuclear emission (hotspots and
lobes), using conservatively large masking regions which effectively
covered the whole extended radio source in each case, and then
generated a radial profile for each object. We chose to represent the
extended emission with isothermal $\beta$ models \citep{cf78}, which
give rise to an angular distribution of counts on the sky of the form
\[
b_{\rm X}(\theta) = b_0 \left(1 + {\theta^2\over\theta_c^2}\right)^{{1\over2} -
3\beta}
\]
where $b_{\rm X}$ is the count density (in count arcsec$^{-2}$) as a function
of off-source angle $\theta$, $b_0$ is the central count density, and
$\theta_c$ and $\beta$ parametrize the angular scale and shape of the
extended emission. Modelling the background-subtracted
radial profiles as a combination of a delta function and a $\beta$
model, both convolved with the point spread function (PSF), we then
carried out least-squares fits using a grid of $\beta$ and $\theta_c$
values to find out whether extended emission was required and to
determine the best-fitting $\beta$ model parameters.

To do this we required an analytical model of the PSF. \citeauthor{wbh01a}
(\citeyear{wbh01a}, \citeyear{wbhl01b}) discussed the fitting of
analytical forms to data from the {\sc ciao} PSF library, and our
approach is similar. Our data are derived from the {\sc ciao} {\sc mkpsf}
command\footnote{This command generates images of the {\it Chandra}
point-spread function at a given energy from the PSF library, which is itself derived
from ray-tracing simulations of the X-ray mirror assembly. See
http://cxc.harvard.edu/ciao/ahelp/mkpsf.html and
http://cxc.harvard.edu/caldb/cxcpsflib.manual.ps for more information.}, energy-weighted to reflect the approximate energy
distribution of the observed data in the extraction radius, smoothed
with a small Gaussian to simulate the effects of pixelation and aspect
uncertainties, and then fitted with a suitable function of radius. The
forms used by \citeauthor{wbh01a}, though a good fit to the inner regions
of the PSF and so adequate for the purposes used in their papers, do
not represent the wings found at radial distances of tens of arcsec,
and these are important in the case of 3C\,263 and 3C\,351, where the
emission is dominated by the bright central point source out to large
radii. Accordingly, we fit a functional form consisting of two
Gaussians, two exponentials and a power-law component, which we found
empirically to be a good fit out to large distances, to the PSF
$a(\theta)$:

\begin{displaymath}
a(\theta) = a_1 \exp\left(-{{\theta^2}\over 2s_1^2}\right) +
       a_2 \exp\left(-{{\theta^2}\over 2s_2^2}\right) +
       a_3 \exp\left(-{\theta\over s_3}\right) + a_4 \exp\left(-{\theta\over s_4}\right) +
       a_5 \left(1 + {\theta\over\theta_n}\right)^{s_5}
\end{displaymath}
where $a_1 .. a_5$, $s_1 .. s_5$ are parameters of the fit and
$\theta_n$ is a fixed normalizing value. In practice, the power-law
index $s_5$ was determined using the lowest-resolution PSF libraries,
and then fixed while the other parameters were determined from
higher-resolution data. In using a power-law model for the PSF wings,
we have followed other work on characterizing the large-radius PSF
(see, e.g., documents at URL:
$<$\url{http://asc.harvard.edu/cal/Hrma/hrma/psf/}$>$) and we have
obtained similar power-law indices, $s_5 \approx 2.0$, in all the fits
we have carried out.

\subsection{Inverse-Compton modeling}

To compare the observed X-ray emission from the lobes and hotspots to
the predictions of inverse-Compton models it is necessary to know the
spatial and spectral structure of these components in the radio and
optical in as much detail as possible. The available data in these
bands are discussed in the following sections, but in
each case we extracted and tabulated radio flux densities (Table
\ref{radiofluxes}) and derived radio-based models for the hotspots and
lobes. Where these components were well resolved, we measured their
sizes directly from radio maps; where they were compact, as was the
case for some of the hotspots, we characterized the component size by
fitting a model of the emission from an optically thin homogeneous
sphere, convolved with the restoring Gaussian, to the
high-resolution radio maps.

We then used a computer code to predict inverse-Compton flux density
as a function of magnetic field strength. Two inverse-Compton codes
were available to us. One, described by \citet{hbw98}, treats all
inverse-Compton sources as homogeneous spheres; this allows us to
neglect the anisotropy of inverse-Compton emission, and so gives a
quick and simple calculation. Our other code is based on the results
of \citet{b00}. By incorporating Brunetti's formulation of anisotropic
inverse-Compton scattering, this code allows us to take account of the
spatial and spectral structure of the resolved hotspots. To model
spatial structure, the hotspots are placed on a fine three-dimensional
grid and the emissivity resulting from the illumination of every cell
by every other cell is calculated (making small corrections for
self-illumination and nearest-neighbour effects). To do this in full
generality is computationally expensive, since it requires numerical
integration of equations A1--5 of \citet{b00} over the electron
distribution and incoming photon distribution for each cell in an
$O(n^6)$ algorithm, where $n$ is the number of cells in one dimension
of the three-dimensional grid. We solve this problem by allowing only
a small number of distinct electron spectra and magnetic fields in our
grid. This means that the task of calculating the illumination of a
region of electrons of a given spectral shape by a given photon
distribution reduces to one of tabulating the integral of Brunetti's
results for a suitably sampled subset of the possible values of the
geometry parameter $k_3$. Since the dependence on the {\it
normalization} of the electron energy spectrum and on the incoming
photon flux is linear, the $O(n^6)$ part of the algorithm then just
involves an interpolation and multiplication. For homogeneous spheres,
or in the case of cosmic microwave background (CMB) inverse-Compton
scattering (CMB/IC) where anisotropy is not an issue, the results of
our two codes agree to within a few per cent, and so in the results
presented below we use whichever code is most appropriate for a
particular situation.

The synchrotron spectra of the hotspots and lobes studied in this work are
not as well studied as those of earlier targets, so some
assumptions are necessary in modeling them. Our basic model is a
broken power-law electron energy spectrum, such that

\[
N(\gamma) = \cases{0&$\gamma<\gamma_{\rm min}$\cr
	N_0 \gamma^{-p}&$\gamma_{\rm min} < \gamma < \gamma_{\rm
	break}$\cr
	N_0 (\gamma_{\rm break}^d) \gamma^{-(p+d)}&$\gamma_{\rm break} < \gamma < \gamma_{\rm
	max}$\cr
	0&$\gamma > \gamma_{\rm max}$\cr
}
\]
where $N(\gamma){\rm d}\gamma$ gives the number density of electrons
with Lorentz factors between $\gamma$ and $\gamma+{\rm d}\gamma$. We
choose this model because synchrotron theory tells us that it is a
good approximation to the expected situation in hotspots, where
particle acceleration and synchrotron losses are competing; for $d=1$
it approximates the standard `continuous injection' model, which
applies to a region containing the acceleration site and a downstream
region of synchrotron radiation loss
\citep[e.g.,][]{hm87}. Theoretical prejudice (based on models of shock
acceleration) also suggests a value $p=2$ for the low-energy power-law
index (the `injection index'). This electron energy spectrum gives
rise to a synchrotron spectrum whose spectral index $\alpha$ is flat
or inverted ($\alpha = -1/3$) around and below some frequency
$\nu_{\rm turn}$, has a value 0.5 up to around some break frequency
$\nu_{\rm break}$ and then steepens to a value of 1.0 before an
exponential cutoff around some frequency $\nu_{\rm cutoff}$. Such
models have been successfully fit to a number of well-studied hotspots
\citep[e.g.,][]{mrhy89,lh00}. It is
well known that the synchrotron turnover, break and cutoff are not
sharp;
for typical parameters of a radio source they occupy a decade or more
in frequency. It is for this reason that we prefer to work with the
well-defined and physically interesting quantities $\gamma_{\rm min}$,
$\gamma_{\rm break}$, $\gamma_{\rm max}$, whose only disadvantage is
that they must be specified together with a value of the magnetic
field strength $B$. We typically work with the equipartition magnetic
field strength, $B_{\rm eq}$, which is given by
\[
m_{\rm e} c^2 \int_{\gamma_{\rm min}}^{\gamma_{\rm max}} \gamma N(\gamma)
{\rm d}\gamma = B_{\rm eq}^2/2\mu_0
\]
(where we have assumed that equipartition is between the radiating
electrons and the magnetic field only, and where SI units are used, with
$\mu_0$ being the permeability of free space).

$\gamma_{\rm max}$ is constrained if there is evidence for spectral
steepening to $\alpha > 1$ in the synchrotron spectrum. We find that
$\gamma_{\rm break}$ is often in the observed radio region, so that it
is relatively easy to constrain. We therefore estimate $\gamma_{\rm
max}$ and $\gamma_{\rm break}$ using least-squares fitting to the
radio and optical data. (Typically $\gamma_{\rm max}$ lies above the
radio region, in which case its value has little effect on our
inverse-Compton calculations.) $\gamma_{\rm min}$, however, can only be
directly constrained by observations of a low-frequency turnover in
sources which are not synchrotron self-absorbed \citep{cpdl91, hbw01a,
h01} or by observations of
optical inverse-Compton emission \citep{h01} and neither of
these methods has yet been applied to the hotspots or lobes of our
sources.  Our low-frequency data on the hotspots typically constrain
$\gamma_{\rm min} \la 1000$ for equipartition fields, but values of
$\gamma_{\rm min}$ between 400 and 1000 have been reported for other
objects \citep[and references therein]{h01}. In addition, it is
possible, as argued by \citet{bbcs02a}, that the electron
spectrum at low energies should be modified by the effects of
adiabatic expansion out of the hotspot. Fortunately, the values of
$\gamma_{\rm min}$ and the low-energy electron spectral shape have
only a small effect on the inverse-Compton calculation, because the
broken synchrotron spectrum produces few high-energy photons that can
be scattered into the X-ray by such low-energy electrons. Unless
otherwise stated we assume $\gamma_{\rm min} = 1000$.

\begin{deluxetable}{lllrr}
\tablewidth{11.2cm}
\tablecaption{Radio flux densities for source components}
\tabletypesize{\scriptsize}
\tablehead{
Source&Component&Model&Frequency&Flux density\\
&&&(GHz)&(mJy)
}
\startdata
3C\,263&Hotspot K&Sphere, $r=0\farcs39$\tablenotemark{a}&1.4&1670\\
&&&4.8&591\\
&&&8.3&303\\
&&&14.9&184\\
&&&84.9&54\\
&Hotspot B&Sphere, $r = 0\farcs 18$&4.8&23\\
&&&8.3&16\\
&&&14.9&9\\[2pt]
&S lobe (X-ray region)&Cube, $9\farcs5 \times 6\farcs3 \times
9\farcs3$&1.4&206\\
&&&4.9&45\\
&N lobe&Sphere, $r=6\arcsec$&1.4&550\\
&&&4.9&188\\
3C\,330&NE hotspot&Cylinder, $2\farcs3 \times 0\farcs64$&1.5&3830\\
&&&8.4&755\\
&&&14.9&376\\
&&&84.9&25\\
&SW hotspot&Sphere, $r = 0\farcs19$&8.4&0.77\\[2pt]
&NE lobe (X-ray region)&Cylinder, $15\arcsec \times 2\farcs$6&1.5&737\\
&&&8.4&161\\
&&&14.9&63\\
&SW lobe (X-ray region)&Cylinder, $15\arcsec \times 2\farcs6$&1.5&768\\
&&&8.4&135\\
&&&14.9&55\\
3C\,351&Hotspot J&Sphere, $r = 0\farcs16$&1.4&530\\
&&&8.4&130\\
&&&15.0&88\\
&&&84.9&17\\
&Hotspot L&Sphere, $r = 0\farcs8$&1.4&1300\\
&&&8.4&316\\
&&&15.0&201\\
&&&84.9&28\\
&Hotspot A&Two spheres, $r = 0\farcs35$&8.4&3\\
&N lobe (X-ray region)&Sphere, $r=10.5\arcsec$&1.4&283\\
&&&8.3&43\\
&S lobe (X-ray region)&Cylinder, $28\arcsec \times 6\farcs5$&1.4&248\\
&&&8.3&46\\
\enddata
\tablenotetext{a}{See the text for a discussion of the more detailed
spatial models applied to this hotspot in the inverse-Compton calculation.}
\tablecomments{Errors on the hotspot flux densities are due primarily
to flux calibration uncertainties, nominally 2\% for the VLA data and
10\% for the 85-GHz BIMA data, as discussed in \S \ref{obs}.}
\label{radiofluxes}
\end{deluxetable}

\section{3C\,263}

\subsection{Introduction}

3C\,263 is a $z=0.66$ quasar. In the radio, the best maps are those of
\citet[hereafter B94]{bhlb94}, who show it to have a one-sided jet
which points towards a bright, compact hotspot in the SE lobe. The
source lies in an optically rich region, and some nearby objects'
associations with the quasar have been spectroscopically confirmed; {\it
HST} observations show several close small companions. Deep X-ray
images were made with {\it ROSAT} by \citet{hegy95} in an attempt to
locate the X-ray emission from the host cluster, but no significant
extended emission on cluster scales was found by them or by
\citet{hw99} who re-analysed their data. The {\it ROSAT} data were
dominated by the quasar's bright, variable nuclear X-ray component.

\subsection{Core}

The bright X-ray nucleus of 3C\,263 suffers from pileup in our
full-frame observation. The piled-up count rate over the full {\it
Chandra} band is $0.190 \pm 0.002$ s$^{-1}$, so that we would expect a
pileup fraction of about 20\%. When the core is fitted with a
power-law spectrum without making any correction for pileup, there is
no sign of any excess absorption over the (small) Galactic value, but
the best-fit power-law is very flat ($\alpha \approx 0.1$) and there
are substantial residuals around 2 keV, which are both characteristic
of piled-up spectra. Using the pileup model in {\sc xspec}, which
implements the work of \citet{d01}, we find a
good fit with a range of steeper power-laws. Following \citet{d01} we
fix the grade correction parameter $\bar g_0$ to 1 and allow only the
grade morphing parameter to vary.\footnote{The model is prone to getting stuck
in local minima, and so we explored parameter space using {\sc
steppar}.} We find the best fit (Table \ref{restab}) to have $\alpha =
0.75 \pm 0.06$ (since the slope and normalization are strongly
correlated in this model, the error quoted is $1\sigma$ for two
interesting parameters), which is in reasonable agreement with
previous observations; for example, in the compilation of \citet{mbc94}
$\alpha$ values between 0.7 and 1.0 are reported. The pileup-corrected
1-keV flux density is substantially lower than the value we estimated
(based on a similar spectral assumption) from the {\it ROSAT} HRI data
\citep{hw99}, suggesting variability by a factor of
$\sim 2$ on a timescale of years (the {\it ROSAT} data were taken in
1993). This is consistent with the observations of \citet{hegy95},
who saw variability by a factor 1.5 in 18 months.

Using the default {\it Chandra} astrometry, the core centroid is at
$11\hh39\mm57\fs170$, $+65\degr47'49\farcs22$, while our best position
for the radio core, using our 15-GHz VLA data, is $11\hh39\mm
57\fs025$, $+65\degr47'49\farcs36$, which is within $0\farcs05$ of the
position quoted by B94. The {\it Chandra} position is
thus offset by approximately $0\farcs89$ in RA and $0\farcs14$ in
declination from the radio data, which is consistent with known {\it
Chandra} astrometric uncertainties. In our spatial analysis we shift
the {\it Chandra} data so that the core positions match, and align
other radio and {\it HST} data with our best core position.

\subsection{Hotspots}

\begin{figure*}
\plotone{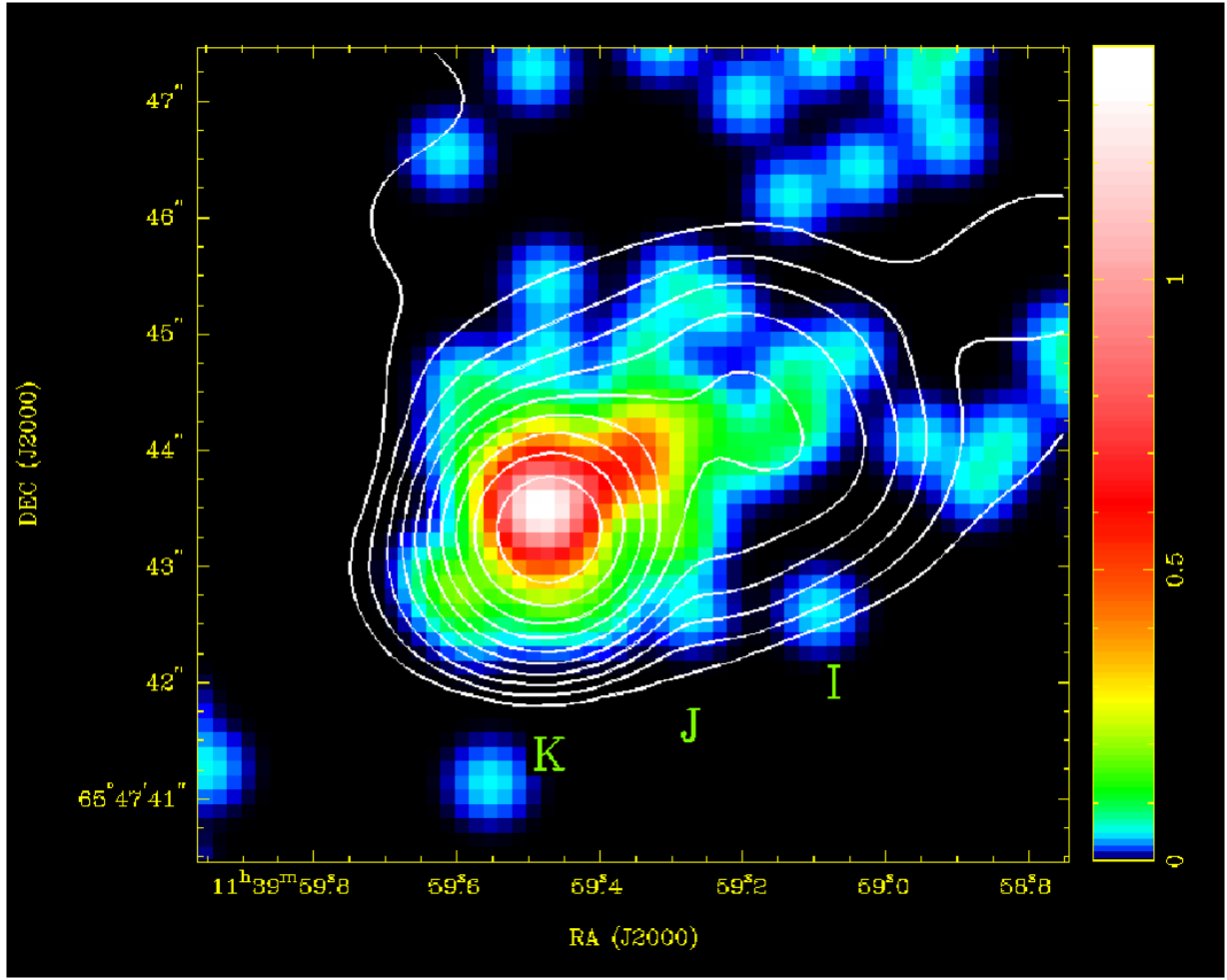}
\caption{The radio and X-ray hotspot of 3C\,263. The X-ray hotspot is
smoothed with a $0\farcs5$ Gaussian. Overlaid are
5-GHz radio contours at a resolution of $0\farcs85$, approximately
matched to the {\it Chandra} data. These are logarithmic in steps of
2, starting at 0.5 mJy beam$^{-1}$. The labels K, H and I refer to
features of the radio emission discussed in the text, following the
notation of B94.}
\label{263-hs}
\end{figure*}

The bright SE hotspot of 3C\,263 is detected in the {\it Chandra}
observation; it corresponds to a faint feature seen in the {\it ROSAT}
HRI images by \citet{hegy95}, which they describe as a `clump' of
X-ray emission. As Fig.\ \ref{263-hs} shows, we detect emission with
{\it Chandra} not just from the compact component K (we use the
notation of B94) but also the faint tail J (which may trace the
incoming jet) and the trailing plateau of emission I. The X-ray/radio
ratio is considerably higher in J (by a factor $\sim 4$) than in the
other components. The high-resolution images show that the X-ray
centroid of the compact component K is slightly to the north of the
peak of the radio emission. The offset is about $0\farcs3$, or 2 kpc.
The angular displacement about the pointing centre, 1.1\degr, is too
large by an order of magnitude to be due to {\it Chandra} roll
uncertainties, which are typically at most 0.1\degr\ (Aldcroft,
private communication); we conclude that the offset is real. We
extracted spectra from the entire complex IJK and from the compact
hotspot K. The net count rate from the IJK complex is consistent,
within the errors, with the count rate estimated by \citeauthor{hegy95} for
the `clump' in their {\it ROSAT} data, converted to an equivalent {\it
Chandra} count rate.

\begin{figure*}
\plotone{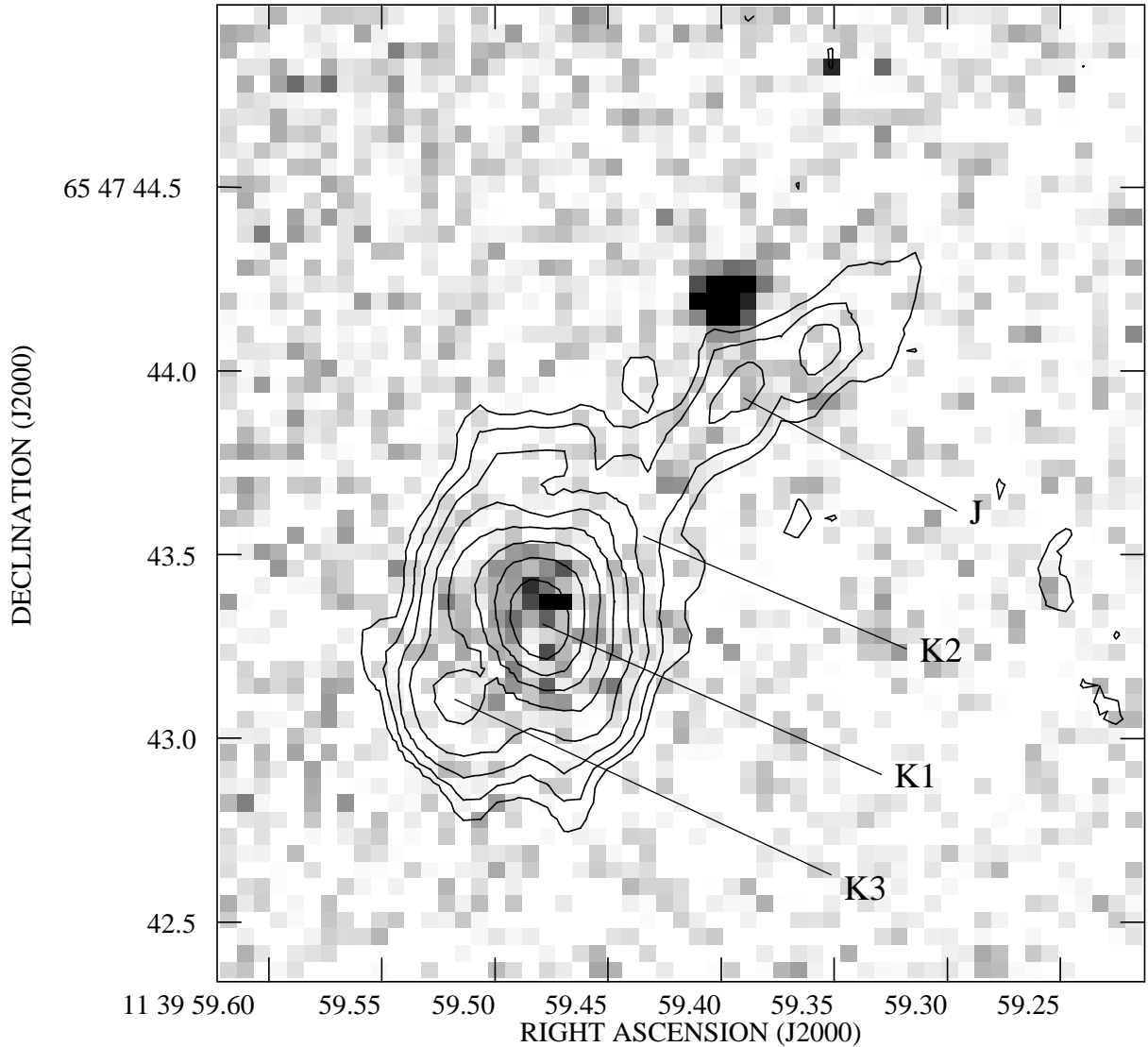}
\caption{Optical hotspot of 3C\,263 with high-resolution radio data. The greyscale runs between 4 and 11 {\it HST} counts per PC pixel.  The radio map is 
the 15-GHz map at full resolution ($0\farcs14 \times 0\farcs10$). The
lowest radio contour is at 0.2 mJy beam$^{-1}$ and the contours
increase logarithmically by a factor 2. The
components K1, K2, K3 of hotspot K, and the linear feature J, discussed in the text, are labeled.}
\label{263HST}
\end{figure*}

A faint optical counterpart to K is detected in the 1000-s F675W {\it
HST} observation of 3C\,263 (Fig.\ \ref{263HST}). After aligning the
radio core with the peak of the optical quasar emission, the peak
optical and radio positions of the hotspot agree to within $0\farcs1$.
({\it HST} roll uncertainties are usually no worse than $0.05\degr$,
corresponding to an error at this distance from the pointing centre of
$0\farcs01$.) We estimate a flux density for this component at $4.5
\times 10^{14}$ Hz of $0.8 \pm 0.4$ $\mu$Jy, with the large error
arising because of uncertainties in background estimation.

\begin{figure*}
\plotone{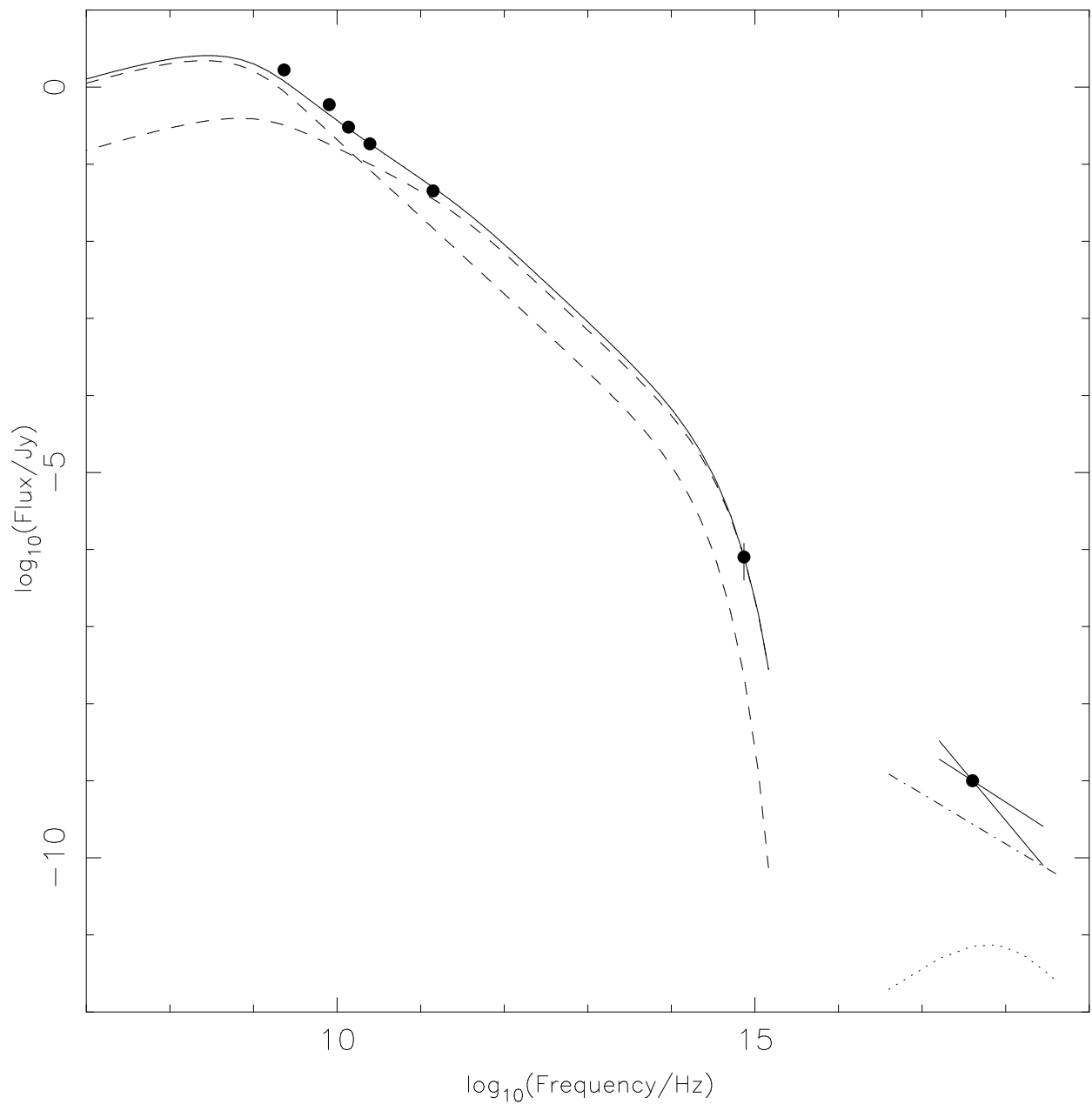}
\caption{The source-frame spectrum of hotspot K of 3C\,263. Radio
points from Table \ref{radiofluxes}, the optical {\it HST} flux
density and the X-ray flux density are plotted together with the
two-component synchrotron model with $\gamma_{\rm
min} = 1000$ discussed in the text (dashed lines show the two
components, solid line shows the total) and the predicted total SSC
emission (dot-dashed line) and CMB/IC emission (dotted line) in this model at equipartition.}
\label{263-sed}
\end{figure*}

As Fig.\ \ref{263HST} shows, the hotspot has considerable spatial
structure in the radio. Even the compact component K is shown by our
15-GHz images to have several sub-components; a bright central
component K1, an extended halo K2, and a weak leading compact
component K3. The radio spectrum of hotspot K (Fig.\ \ref{263-sed}),
which we obtain by Gaussian fitting to the 1.4-GHz map and by
integration of matched regions on the higher-resolution 4.8, 8.3 and
15-GHz maps, also shows signs of multiple structure, in that the
spectrum steepens between 4.8 and 8.3 GHz but then flattens again
between 8.3 and 15 GHz. 8.4--15 GHz spectral index maps show that the
compact central component K1 has a flatter two-point radio spectrum,
$\alpha \approx 0.8$, while the surrounding material is
steeper-spectrum, $\alpha \approx 1.0$. Correcting for background
contamination, $\alpha_{\rm K1} \approx 0.7$ between 8.3 and 15 GHz,
which is consistent with the observed 15-85 GHz spectral index. We can
crudely model hotspot K as being a superposition of two of our
standard spectra with different break frequencies, with the K1
component breaking above 85 GHz ($\gamma_{\rm break} \approx 2 \times
10^4$, assuming equipartition magnetic fields) while the K2/3
components break at 1.4 GHz or lower ($\gamma_{\rm break} \approx
1000$). In order for the K1 component not to overproduce the observed
optical emission, assumed to be synchrotron emission, its spectrum
must then steepen further at high energies; we choose to model this
with a high-energy cutoff in the electron energy spectrum,
$\gamma_{\rm max} = 5 \times 10^{5}$. For simplicity, we assign the
extended component the same high-energy cutoff, which makes its
contribution to the observed optical emission negligible. We modeled
K1 as a sphere of radius $0\farcs15$ and K2 as a sphere of radius
$0\farcs39$, based on the 15-GHz map. This synchrotron model is
plotted in Fig. \ref{263-sed}).

Three important unknown factors remain: the geometry of K1/K2 (does
the compact component K1 lie inside or outside, in front of or behind,
the more extended component K2?), the spatial electron distribution in
the two spheres (uniform or centrally peaked? the radio data are not
good enough to say) and the low-energy cutoff in the synchrotron
spectrum. The 1.4-GHz observations suggest a low-energy cutoff
$\gamma_{\rm min} \la 1000$; a higher cutoff would mean we would start
to observe spectral flattening at this frequency. We carried out our
inverse-Compton modeling with a range of possible geometries, spatial
electron distributions and $\gamma_{\rm min}$ values in order to
assess the effect of these differences. We find that the most
significant effect is given by a more centrally peaked distribution of
electrons; $\gamma_{\rm min}$ has little effect, and only if K1 and K2
are widely separated is a significant difference made by the geometry
(arising because of the weaker mutual illumination of K1 and K2). The
overall conclusion from this modeling is that for equipartition
magnetic fields (where equipartition in this case means that
equipartition holds separately for each spectral component) the
inverse-Compton flux density from hotspot K is typically less than the
observed value by a factor $\sim 3.5$. To produce the observed X-ray
flux density by the inverse-Compton process, the magnetic field
strengths in the two components of the hotspot must be lower than the
equipartition value by a factor 1.9 (for the case where the compact
component is at the center of the extended component), so that the
compact component has $B = 15$ nT and the extended component has $B=7$
nT. The predicted X-ray spectral index at an observed energy of 1 keV
is then 0.75, which is consistent within the $1\sigma$ error with the
observed value, $1.0 \pm 0.3$. This conclusion is robust to changes in
the model parameters; for example, even a simple homogeneous spherical
model, while giving different values for the equipartition magnetic
field strength, predicts a net inverse-Compton flux density of 0.26
nJy and requires a factor $\sim 2$ decrease in field to reproduce the
observed X-rays.

The comparatively strong detection of component J is puzzling. Its
1-keV flux density is $\sim 0.35 \pm 0.1$ nJy, assuming spectral
parameters similar to those of K. If the X-ray emission corresponds to
the narrow feature seen in Fig. \ref{263HST}, it is much too bright to
be SSC at or close to equipartition. A long, thin cylinder is a very
poor SSC source, and the equipartition 1-keV flux density of this
component should be $\sim 2$ pJy, including scattering of photons from
K1 and K2. The magnetic field would have to be lower than
equipartition by a factor $\sim 20$ if all the X-ray flux were to be
produced by component J by the SSC and invese-Compton processes. If
there is significant relativistic motion in J, then in its frame the
emission from K and from the CMB is enhanced, and this can increase
the net X-ray emissivity; however, it is hard to achieve the required
increase without very large bulk Lorentz factors and small angles to
the line of sight. Our model here is unrealistically simple, as it
does not take acount of the extended emission around J: however, it
seems clear that this region is significantly different from hotspot
K.

The much fainter NW compact hotspot, B in the notation of B94, is not
detected in X-rays. Assuming that the source is unresolved to {\it
Chandra} and using a detection cell of 6 standard {\it Chandra} pixels
(the 80\% encircled energy criterion, as used by {\sc celldetect}), we
can place a $3\sigma$ upper limit on its count rate of $1 \times
10^{-4}$ s$^{-1}$ over the 0.4-7.0 keV energy band, estimating the
background count level from the region around the hotspot, and convert
this to a limit on flux density by using the spectrum of the detected
hotspot. No optical counterpart to this hotspot is visible in the {\it
HST} image. Its predicted inverse-Compton X-ray flux density, assuming
similar electron spectral parameters to those used for hotspot K, a
homogenous spherical model, and an equipartition magnetic field, is 5
pJy, an order of magnitude below our upper limit. We can say only that
$B$ cannot be more than a factor $\sim 4$ below $B_{\rm eq}$
in this hotspot.

\subsection{Jet}

There is no evidence of X-ray emission from 3C\,263's jet in the data,
with the possible exception of the hotspot-related feature J,
discussed above. The jet is $11\farcs6$ long in the radio images of
B94 and is essentially unresolved transversely at their highest
resolution of $0\farcs35$. Using a background region in the SE lobe,
we estimate a $3\sigma$ upper limit on its count rate, as tabulated in
Table \ref{restab}. We convert this to an upper limit on the jet X-ray
flux density by using the spectrum of the detected hotspot.

The radio flux density of the jet is only 8 mJy at 5 GHz, so that its
radio to X-ray spectral index is $\ga 1.0$. If the X-ray emission from
quasar jets is boosted inverse-Compton scattering from the CMB, as
suggested by various authors \citep[e.g.,][]{tmsu00}, then its
expected X-ray flux density depends on the bulk Lorentz factor
$\Gamma$ and the angle made by the jet to the line of sight $\theta$. We
find the upper limit to be inconsistent only with extreme models of
the jet, with $\Gamma \ga 7$ and $\theta \la 10\degr$. The overall
appearance of the source suggests that the angle to the line of sight
is a good deal larger than this, in which case even larger bulk
Lorentz factors are not ruled out by the X-ray data.

\subsection{Lobes}

\begin{figure*}
\plotone{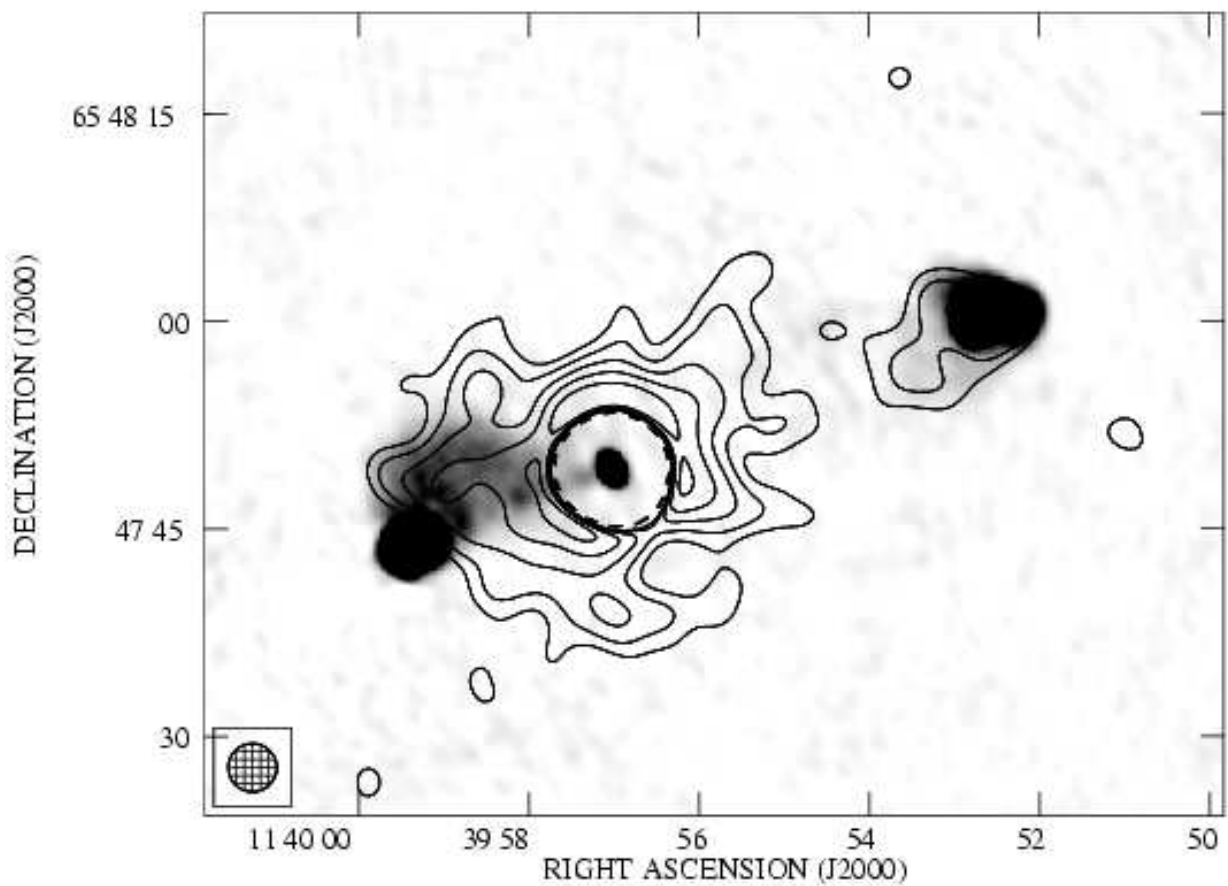}
\caption{The lobes and extended emission of 3C\,263. The contours show
the X-ray emission smoothed with a $3\farcs5$ (FWHM) Gaussian. The
hotspot regions were masked out before smoothing and a scaled,
smoothed, energy-weighted PSF model was subtracted from the image
after smoothing to remove the nuclear emission. Because the X-ray core
of 3C\,263 is piled up, this subtraction leaves some residuals at the
center of the smoothed image. The lowest contour is the $3\sigma$
level, defined according to the prescription of \citet{h00}, and
the contours increase logarithmically by a factor $\sqrt{2}$. The FWHM
of the convolving Gaussian is shown in the bottom left-hand corner.
The greyscale shows a 1.4-GHz VLA image with $1\farcs5 \times
1\farcs1$ resolution; black is 10 mJy beam$^{-1}$.}
\label{263lobe}
\end{figure*}

X-ray emission is detected from the lobes of 3C\,263 (Fig.\
\ref{263lobe}). Although the detection of the NW lobe emission is more
obvious, because it is further from the core and there is no compact
hotspot emission, we also detect excess counts from the SE lobe.

For inverse-Compton calculations we model both lobes with similar
broken power-law electron distributions. We take $\gamma_{\rm min}$
here to be 100 rather than 1000, to take some account of the effects
of adiabatic expansion; the inverse-Compton prediction is insensitive
to this choice. $\gamma_{\rm break}$ is required to correspond to a
frequency near the radio region, to produce the steep spectrum of the
lobes, and we model both lobes with high-energy cutoffs in the radio
region, $\gamma_{\rm max} \sim 2 \times 10^4$. The predicted 1-keV
CMB/IC flux densities on this
model (which treats the lobes as uniform) are 0.13 nJy for the S lobe
and 0.2 nJy for the N lobe, a factor $\sim 4$ below the observed flux
densities in both cases. The magnetic field strength in the lobes must
be a factor $\sim 2$ below the equipartition value if the lobe X-ray
emission is to be produced by inverse-Compton scattering of the CMB.
With this reduction in field strength, the pressures in the lobes due
to electrons and magnetic field are between 1.5 and $2.0 \times
10^{-12}$ Pa.
\label{263lobep}
\subsection{Extended emission}

\begin{figure*}
\plotone{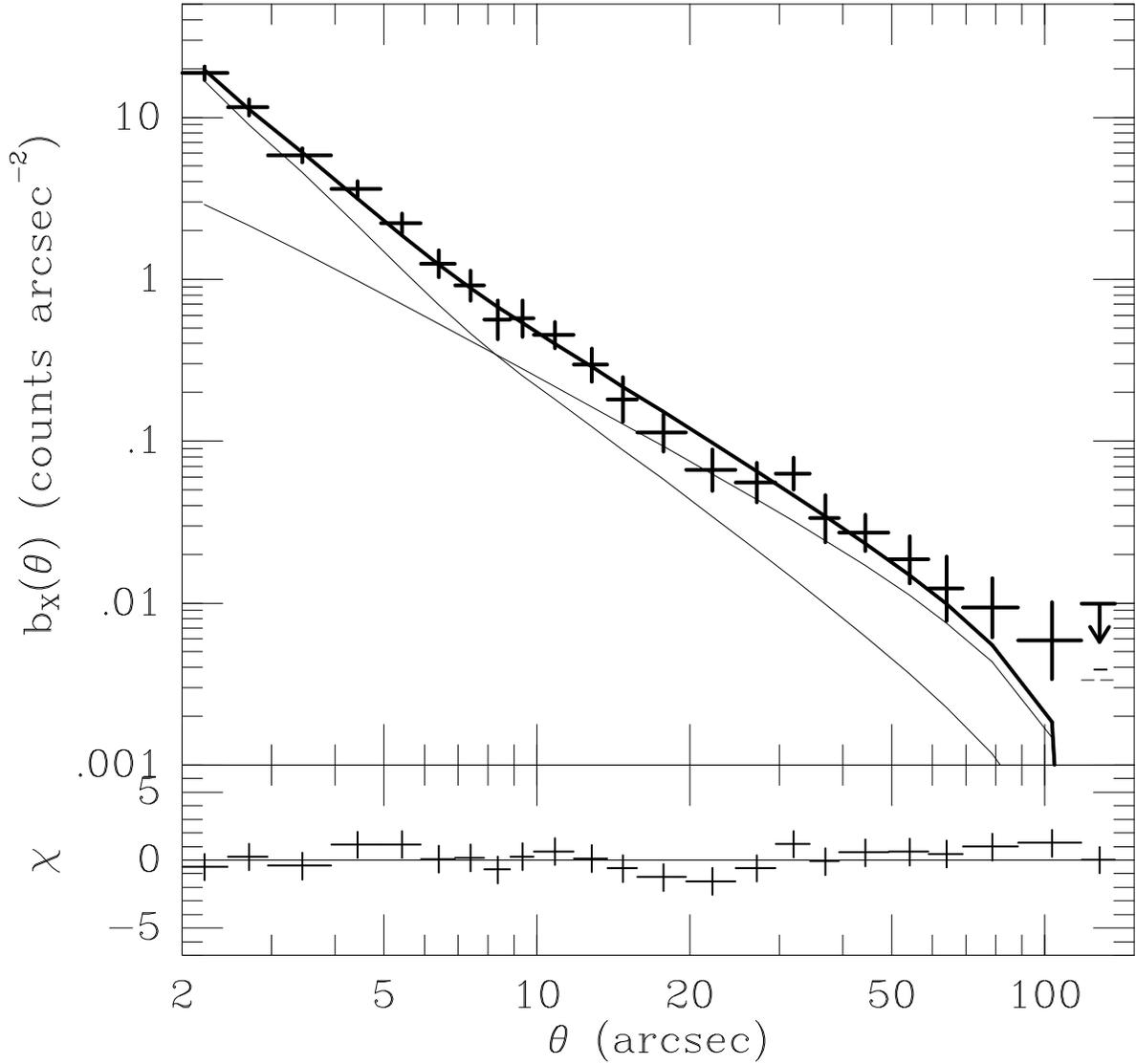}
\caption{Background-subtracted radial profile of 3C\,263 after
hotspot- and lobe-related emission has been masked out. The two light
solid lines show the PSF (steeper slope) and extended model (flatter
slope) fitted to the source, and the heavy line is their sum.}
\label{263profile}
\end{figure*}

3C\,263 shows evidence for extension above the wings of the PSF out to
the edge of the S3 chip (Fig.\ \ref{263profile}). There are
$823_{-120}^{+64}$ net 0.5--7.0-keV counts in the extended component,
and the best fitting model has $\beta = 0.45_{-0.08}^{+0.07}$,
$\theta_c = 0\farcs9_{-0.9}^{+4}$, where errors are $1\sigma$ for two
interesting parameters: no lower limit is set on the core radius, so
that the models are effectively power-law models outside the central
regions.  The background level is too high to allow us to extract a
useful spectrum of the extended emission region, but assuming a
temperature of a few keV, the count rate implies a bolometric X-ray
luminosity in the region, which extends out to 120 arcsec (0.9 Mpc),
of $2$ -- $3 \times 10^{44}$ ergs s$^{-1}$, consistent with the {\it
ROSAT}-derived upper limits and with the presence of a
spectroscopically confirmed cluster around 3C\,263. This luminosity
would correspond, on the temperature-luminosity relation for clusters
\citep[e.g.,][]{wxf99} to a temperature $kT \approx 3$ keV. Using the
analysis of \cite{bw93} to convert the parameters of the $\beta$-model
to physical conditions in the extended hot gas, we find that the
central density on the best-fitting model would be $2 \times 10^5$
m$^{-3}$ and the central pressure $1.9^{+7.0}_{-1.7} \times 10^{-10}$
Pa (errors are determined from the $\beta$-model fits and do not
include temperature uncertainties). The pressure at the distance of
the lobes is better constrained: at $20\arcsec$ (150 kpc) it is
$2.8^{+0.3}_{-0.4} \times 10^{-12}$ Pa for the adopted temperature. If
the source is at a small angle to the line of sight, the external
pressures corresponding to the lobes will be lower. These pressures
are similar to the internal pressures of the lobes determined above,
\S\ref{263lobep}.

\section{3C\,330}

\subsection{Introduction}

3C\,330 is a narrow-line radio galaxy with $z=0.55$. Radio images of
the hotspots are presented by \citet*{fbp97} and by
\citet[hereafter G02]{grpa02}. The lobes are best seen in
lower-resolution maps by \citet{lms89}. Optical
clustering estimates \citep{hl91} imply a reasonably rich
environment for 3C\,330, but no X-ray emission was detected from the
source in the off-axis {\it ROSAT} observations discussed by
\citet{hw99}. The observations we report here are thus
the first X-ray detection of 3C\,330.

\subsection{Core}

\begin{figure*}
\plotone{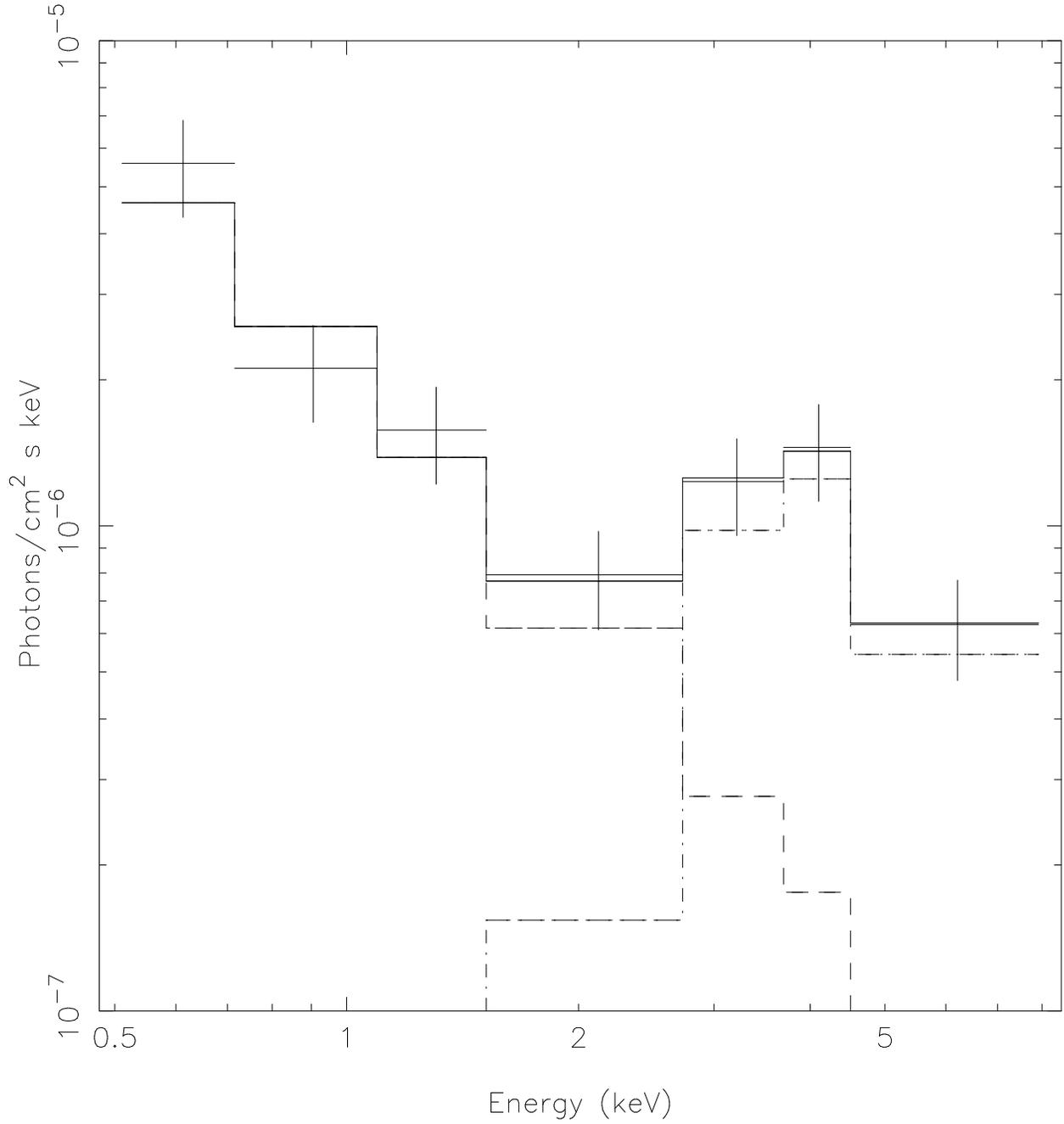}
\caption{X-ray spectrum of the core of 3C\,330.
Solid crosses show the X-ray data. The lines show the model discussed
in the text: the dashed line is the unabsorbed power law, the
dot-dashed line is the absorbed component, and the solid line is the
total model.}
\label{330spec}
\end{figure*}

3C\,330's core has a relatively complex X-ray spectrum (Fig.\
\ref{330spec}). Unlike the cores of other FRII radio galaxies that we
have studied with {\it Chandra} \citep{wbhl01b,hbw01a}, it is not
adequately fitted with a simple absorbed power-law model. The simplest
model that gives a good fit consists of two power laws, one with
Galactic absorption and one with an additional absorption column,
intrinsic to the radio source, of $(4 \pm 2) \times 10^{23}$
cm$^{-2}$. This column density is comparable to that inferred from
hard X-ray observations in Cygnus A \citep{ukny94}, although the
errors are large. Such high column densities are conventionally
explained as being a signature of the dense, dusty torus which is
invoked in unified models to obscure the quasar nucleus and broad-line
regions in narrow-line radio galaxies. In previous work \citep[and
references therein]{hw99} we have argued that soft X-ray
emission can arise in a component related to the radio core, through
inverse-Compton or synchrotron emission: this component can originate
on scales larger than those of the torus and so is not heavily
absorbed. In 3C\,330, it seems plausible that the unabsorbed power law
is this radio-related component, while the heavily absorbed component
is due to the hidden quasar. 3C\,330's radio core is comparatively
weak (only 0.7 mJy at 5 GHz: \citealt{fbp97}), which may explain
why we are able to see the heavily absorbed component in this source
but not in others, where a stronger radio-related component
dominates. If we remove the absorbing column, the (observer frame)
2--10 keV flux of the absorbed component is $\sim 10^{-13}$ ergs
cm$^{-2}$ s$^{-1}$ (with a large uncertainty due to the poorly
constrained spectrum). This is only an order of magnitude less than
the fluxes in the same band that we determine for the two quasars in
our sample, implying luminosities which are not very dissimilar, given
the similar redshifts. The 1-keV flux density for the unabsorbed
component is in good agreement with the radio/soft-X-ray correlation
of \citet{hw99}, and is consistent with the {\it
ROSAT} upper limit on 3C\,330's 1-keV flux density presented there.

Because both the radio and the X-ray cores are fainter in this source
than in the other sources in our sample, the radio-X-ray alignment is
less certain. The X-ray centroid, using the default {\it Chandra}
astrometry, is at $16\hh09\mm34\fs919$, $+65\degr56'37\farcs37$, while
our best radio position (from the map of G02) is
$16\hh09\mm34\fs896$, $+65\degr56'37\farcs69$, in close agreement with
the position found by \cite{fbp97}. This implies a
radio-X-ray offset of $0\farcs35$, and we correct the X-ray data
accordingly.

\subsection{Hotspots}

\begin{figure*}
\plottwo{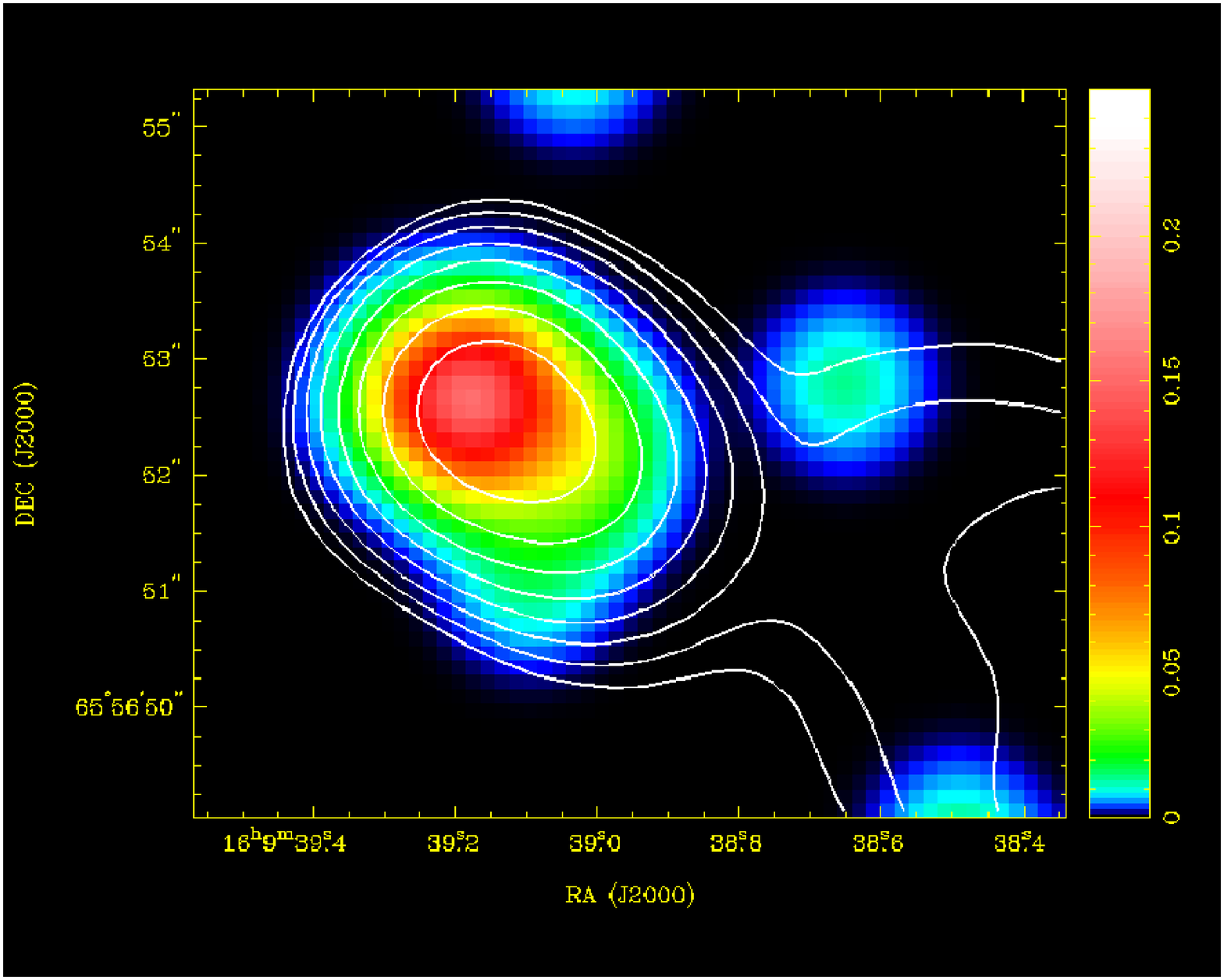}{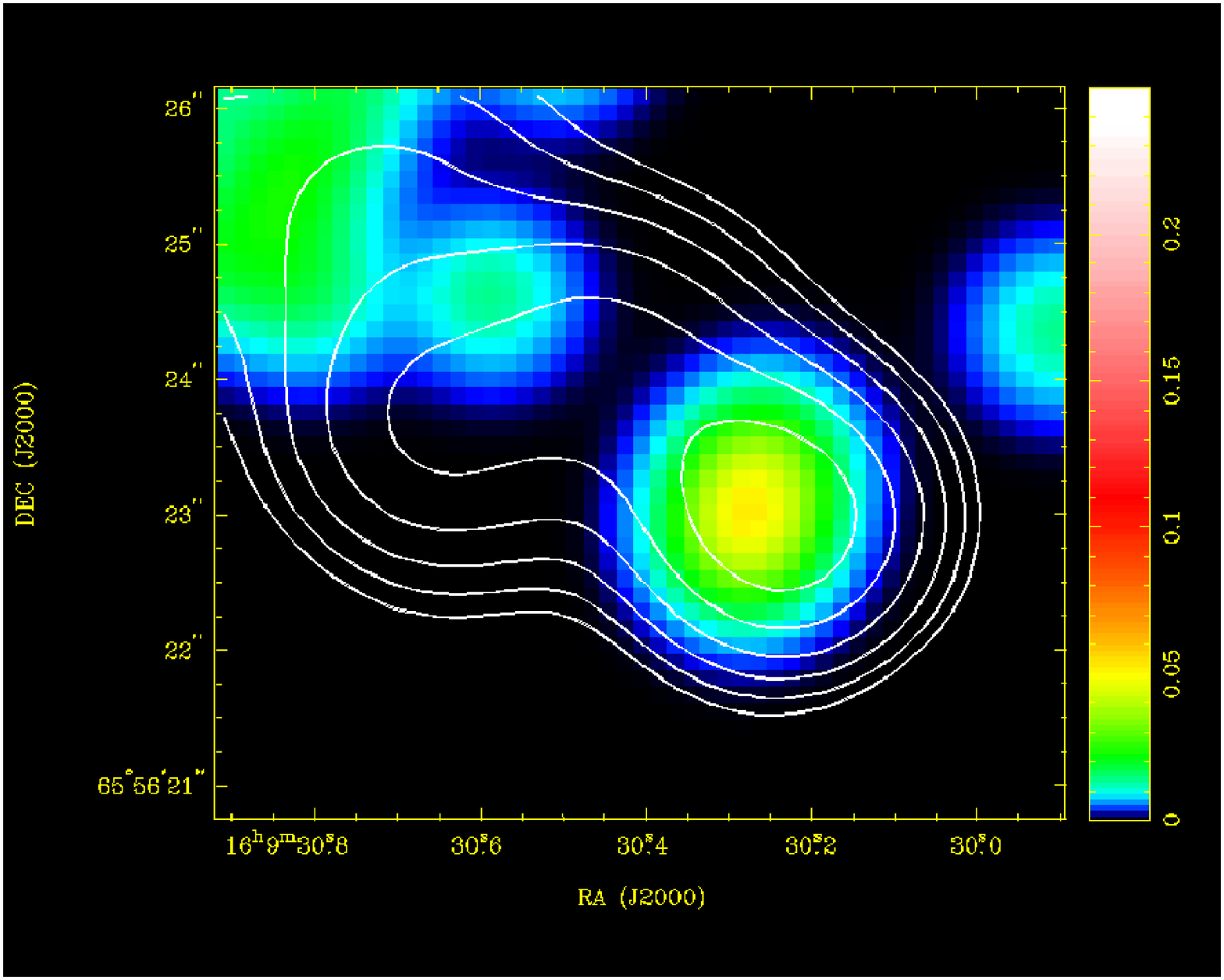}
\caption{
The X-ray hotspots of 3C\,330 smoothed with a 1\arcsec\ Gaussian.
Contours in both cases are from an 8.4-GHz radio map with a resolution of
$1\farcs2$, approximately matched to the X-ray resolution. The
contours increase logarithmically by a factor 2 and the lowest contour
is at 1 mJy beam$^{-1}$. Left: NE hotspot; Right: SW hotspot.}
\label{330hs}
\end{figure*}

Fig.\ \ref{330hs} shows that the NE radio-bright hotspot of 3C\,330 is
clearly detected in the X-ray. There is also a weak but significant
detection of the SW hotspot. The X-ray peak of the brighter hotspot
agrees well with the radio flux peak, and it may be marginally
extended along the same axis as the bright radio-emitting region.

There are insufficient net counts even in the brighter hotspot to
extract a spectrum. We convert the count rates to 1 keV flux densities on
the basis of a power-law spectrum with $\alpha_{\rm X} = 0.5$ and Galactic
absorption. The conversion factor is not sensitive to the precise
choice of spectral index.

Neither hotspot is detected in the {\it HST} observation. We set a
$3\sigma$ upper limit on each hotspot's optical flux density of 0.5
$\mu$Jy at $5.5 \times 10^{14}$ Hz.

\begin{figure*}
\plotone{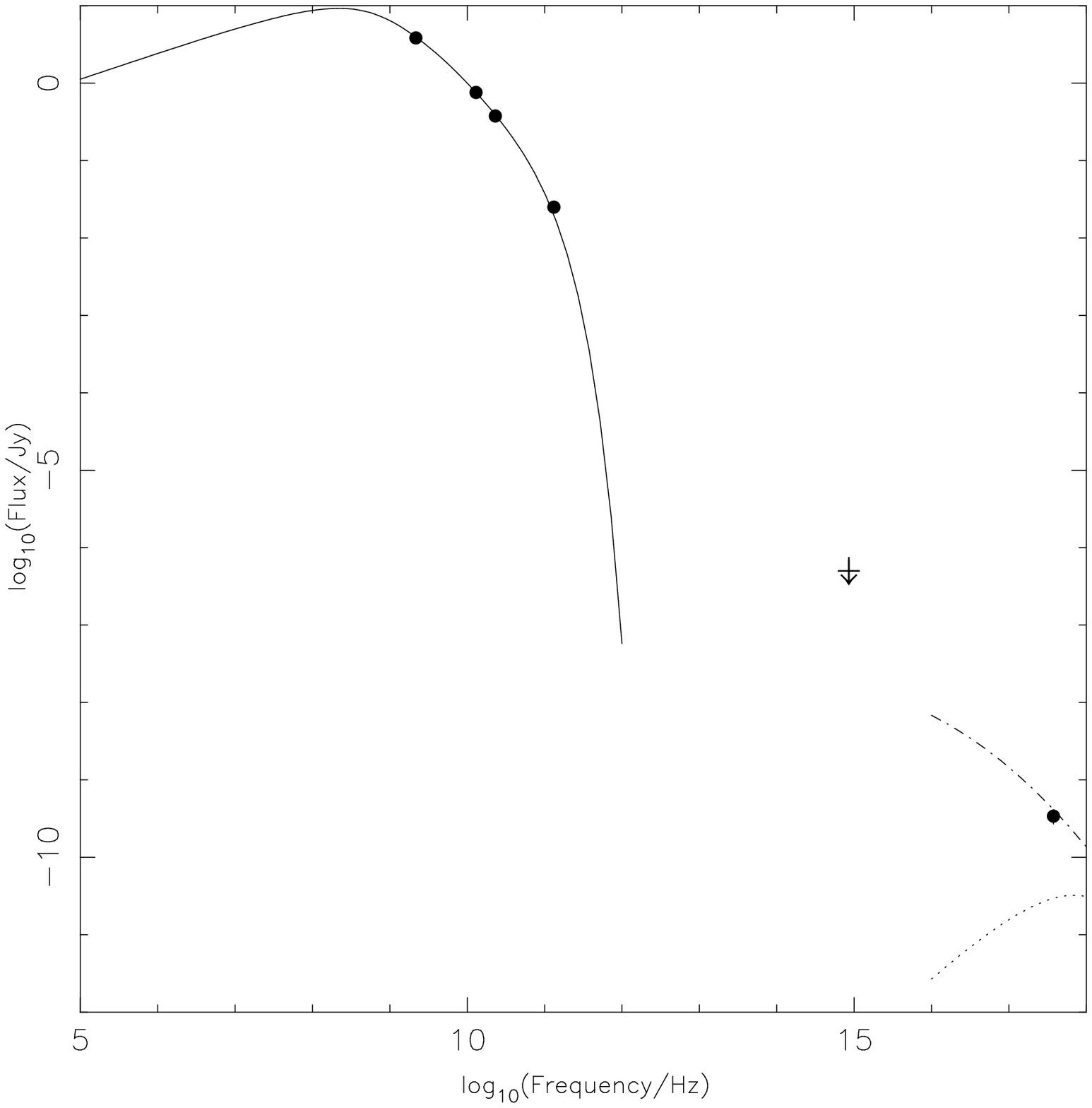}
\caption{The source-frame spectrum of the N hotspot of 3C\,330. Radio
points from Table \ref{radiofluxes}, the optical {\it HST} flux
density limit and the X-ray flux density are plotted together with the
best-fitting synchrotron model with $\gamma_{\rm min} = 1000$ (solid
line) and the predicted SSC (dot-dashed line) and CMB/IC (dotted line)
emission from the hotspot at equipartition. The high-energy cutoff is set to
$\gamma_{\rm max} = 1.4 \times 10^4$ in this figure.}
\label{330-sed}
\end{figure*}

The spectrum of the NE hotspot of 3C\,330 is
plotted in Fig.\ \ref{330-sed}. The structure of this hotspot is quite
well resolved by the high-resolution image of G02; it can be modeled
adequately as a cylinder with length $2\farcs3$ and radius $0\farcs64$
with a linear intensity gradient along its length (which we model as a
linear increase in electron spectral normalization).

The spectrum of the hotspot at radio frequencies is steep
($\alpha\approx 1.0$), and, since we only have one high-resolution map
and cannot resolve any spectral structure of the hotspot, we simply
model it as a single broken power-law model with the break between
spectral indices of 0.5 and 1.0 occurring close to 1.4 GHz. For
equipartition, $\gamma_{\rm break} = 3000$. We set $\gamma_{\rm min} =
1000$. The best fit to the radio observations, including the
high-frequency BIMA data, is well fit with a model with $\gamma_{\rm
max} = 1.4 \times 10^4$; the spectral index between 15 and 85 GHz is
steeper than 1.0, which requires the cutoff to be at low energies. In
this model, the equipartition field strength is 9.5 nT, and the
predicted SSC flux density is 0.49 nJy. If the hotspot is not in the
plane of the sky, projection means that the actual long axis of the
hotspot is longer than we have assumed, and the predicted SSC
emissivity is reduced, by about 10\% for an angle to the line of sight
of 45 degrees. Given these uncertainties, we can say that the observed
SSC emission in the NE hotspot, with a 1-keV flux density of $0.34 \pm
0.07$ nJy on simple spectral assumptions, implies a magnetic field
equal to or slightly {\it higher than} the equipartition value.

The compact component of the SW hotspot is only resolved from the
others at 8 GHz, so we cannot measure its spectrum; we assume the same
electron spectral values as for the NE hotspot, with a spherical
model. The predicted 1-keV SSC flux density from this component at
equipartition ($B = 18$ nT) is then between 0.03 and 0.04 nJy,
depending on the value of $\gamma_{\rm max}$. This is consistent with
the observed flux density of $0.09 \pm 0.04$ nJy, given the large
uncertainties, but the observations are better fit with a magnetic
field strength a factor $\sim 1.5$ below equipartition.

\subsection{Lobes}

\begin{figure*}
\epsscale{0.8}
\plotone{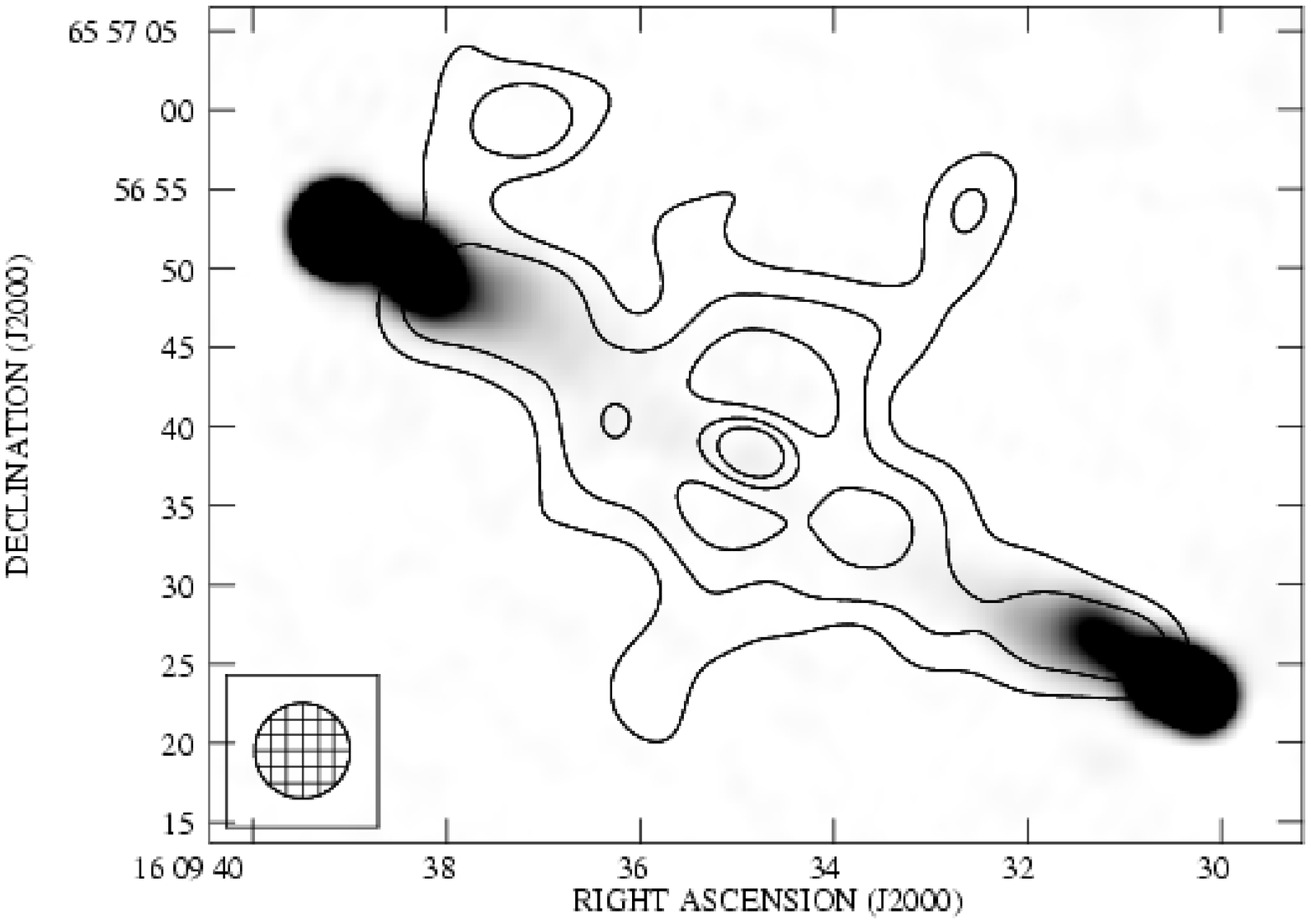}
\caption{The lobes and extended emission of 3C\,330. The contours show
the X-ray emission smoothed with a 6\arcsec\ (FWHM) Gaussian. The
hotspot regions were masked out before smoothing and a scaled,
smoothed, energy-weighted PSF model was subtracted from the image
after smoothing to remove the nuclear emission. The lowest contour is
the $3\sigma$ level, defined according to the prescription of
\citet{h00}, and the contours increase logarithmically by a
factor $\sqrt{2}$. The FHWM of the Gaussian used to convolve the X-ray
emission is shown in the bottom left-hand corner. The object to the NW
of the NE hotspot seems to be a faint background X-ray source. The
greyscale shows the 8.4-GHz VLA image with 3\arcsec\ resolution of
G02; black is 10 mJy beam$^{-1}$. }
\label{330lobe}
\end{figure*}

3C\,330's lobes are both clearly detected in the X-ray (Fig.\
\ref{330lobe}). We defined rectangular extraction regions which avoid
the core (starting at about 10\arcsec\ away from it) and the hotspots, using
parallel adjacent rectangles on either side to give a local background
subtraction. The lobes each contain about 20 net counts in the
0.4--7.0 keV range, so it is not possible to extract useful spectra. As
with the hotspots, we convert the count rate to 1-keV flux density on the
basis of a power law with $\alpha_{\rm X} = 0.5$ and Galactic absorption.

We model the lobes with the same spectral assumptions as for 3C\,263
(\S\ref{263lobep}). In this case the predicted 1-keV flux densities
from CMB/IC at equipartition are 0.17 and 0.18 nJy. Given the large
errors on the measured lobe fluxes, we cannot rule out equipartition
in this case; taking the measured fluxes at face value, they imply
magnetic field strengths a factor $\sim 1.4$ below equipartition.
These would imply internal lobe pressures from electrons and magnetic
field of around $1.5 \times 10^{-12}$ Pa.

\subsection{Extended emission}

\begin{figure*}
\plotone{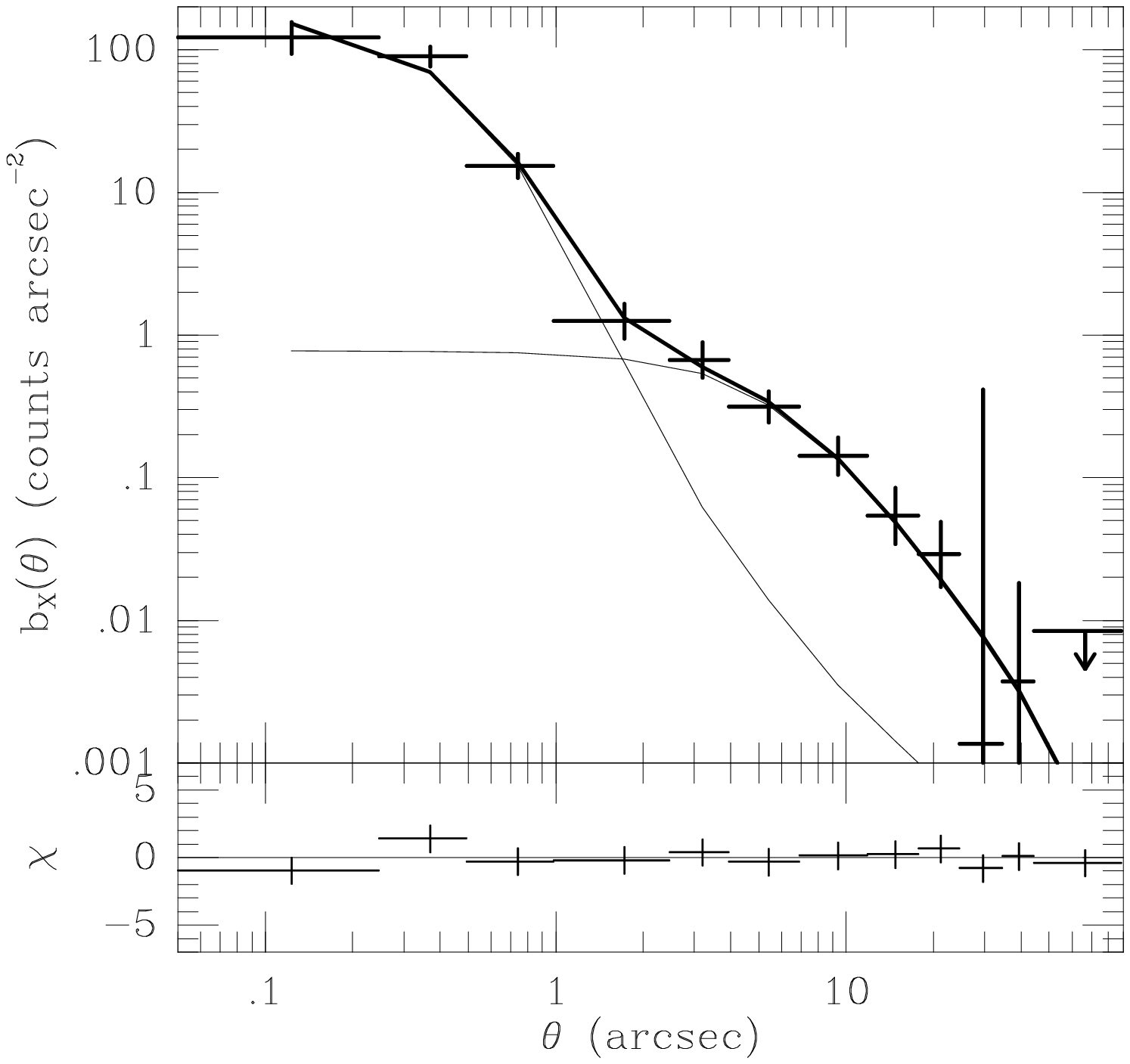}
\caption{Background-subtracted radial profile of 3C\,330 after
hotspot- and lobe-related emission has been masked out. The two light
solid lines show the PSF (smaller scale) and extended model (larger
scale) fitted to the source, and the heavy line is their sum.}
\label{330profile}
\end{figure*}

3C\,330 has weak extended emission (Fig.\ \ref{330profile}), with the
fits showing evidence for only $186_{-37}^{+44}$ extended counts out
to 70 arcsec (0.5 Mpc). The best-fitting $\beta$ model has $\beta =
0.65_{-0.18}^{+1.3}$, $\theta_c = 5\farcs9_{-3.7}^{+15}$ (errors are
$1\sigma$ for two interesting parameters). The luminosity of this
extended component is then around $3 \times 10^{43}$ ergs s$^{-1}$,
corresponding to a temperature around $1.5$ keV. In the best-fitting
model, the central density is $\sim 2 \times 10^4$ m$^{-3}$ and the
central pressure $8.4^{+9.1}_{-3.8} \times 10^{-12}$ Pa. The pressure
at distances corresponding to the lobes, assuming little projection of
this source, varies from $2.2^{+0.4}_{-0.3} \times 10^{-12}$ Pa (at
$10\arcsec$, 70 kpc) to $3.3^{+1.0}_{-2.0} \times 10^{-13}$ Pa (at
$30\arcsec$, 210 kpc). These are similar to the estimate of the
internal lobe pressure above: the lobes would be in pressure balance
at a radius of around $10\arcsec$ but could be overpressured with
respect to the external medium at larger distances.
\label{330ext}

\section{3C\,351}

\subsection{Introduction}

3C\,351 is a well-studied $z=0.37$ quasar. Radio images have been
presented by \cite{lp91}, B94, and G02; their most striking
feature is the bright double hotspot pair to the N and the displaced
nature of the N lobe. The hotspots are also detected in the optical
\citep{r89, lv99}. The X-ray emission from the bright
nucleus has been extensively studied \citep{femw93,mwef94,nfpe99} and it has been argued that the unusual
X-ray spectrum seen in {\it ROSAT} PSPC data is due to an ionized
absorber, often seen in Seyfert 1 galaxies but a rare feature in
quasars. No extended X-ray emission was detected with {\it ROSAT}
\citep{hw99}, but the hotspots were seen in a short
{\it Chandra} observation \citep{bbcp01}.

\subsection{Core}

\begin{figure*}
\plottwo{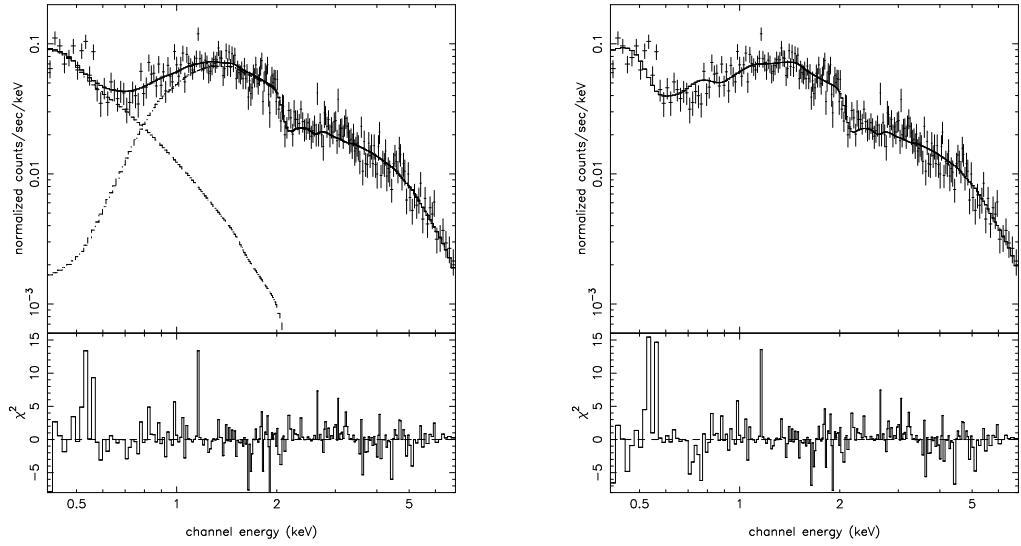}{f13b.eps}
\caption{X-ray spectrum of 3C\,351 nucleus. Left: the spectrum with the best-fitting two-power-law model.
Right: the spectrum with the best-fitting ionized-absorber model. The
lower panels show the contributions to $\chi^2$, plotted with the sign
of the residual.}
\label{351core}
\end{figure*}

The X-ray centroid, using the default {\it Chandra} astrometry, is at
$17\hh04\mm41\fs326$, $+60\degr44'30\farcs22$. 3C\,351's radio core
contains two compact components of similar 8-GHz flux density
separated by $0\farcs35$. B94 and G02 both argue that the southern
component is the true core, while the northern component is a jet
knot. The radio position of the southern component, using the 8-GHz
radio map of G02, is $17\hh04\mm41\fs364$, $+60\degr44'30\farcs46$.
This is in good agreement with the position given by B94 and implies a
radio-X-ray misalignment of $0\farcs37$. We shift the X-ray data to
align the X-ray core with the southern radio component. However, it is
possible that some of the X-ray emission observed from the core region
comes from the northern radio component, since the two components
would not be well resolved by {\it Chandra}. We see no evidence for
extended emission from the X-ray core, and no difference between the
core centroids in different X-ray energy ranges, but the possibility
means that the alignment of the radio and X-ray frames is less secure
than it would otherwise be.

We obtain many counts in the X-ray core, and so detailed spectroscopic
analysis is possible. Unsurprisingly, in view of previous work, a
single power-law model with Galactic absorption is an unacceptable fit
to our data ($\chi^2/n = 638/215$, where $n$ denotes degrees of
freedom); we see strong residuals at 0.6--0.8 keV. A two-power-law
model, with one of the power-law components having additional
intrinsic absorption (as used above for 3C\,330) is a better fit,
with $\chi^2/n = 304/212$ and an intrinsic cold absorbing column of
$(1.1 \pm 0.1) \times 10^{22}$ cm$^{-2}$; but the best-fitting
spectral index of the unabsorbed component in this model is well
constrained and steep, $\alpha = 2.5 \pm 0.3$. The absorbed power law
has $\alpha = 0.46 \pm 0.05$. This model is plotted in Fig.\
\ref{351core}. If the absorbed component is associated with the active
nucleus and the unabsorbed component with the radio emission, then the
absorbing column, which represents only 2 magnitudes of visual
extinction, is small enough to be compatible with the
predictions for a quasar within unified models, while the 1-keV flux
density of the unabsorbed component would be 14 nJy, which is
consistent with the radio-X-ray relation of \citet{hw99}.

We fit two alternative models, consisting of a single power law with
Galactic absorption plus either `windowed' absorption (the {\sc
zwndabs} model in {\sc xspec}), which roughly mimics a warm absorber,
or the {\sc xspec} ionized absorber model ({\sc absori}). The {\sc
zwndabs} model gives a less good fit than the power-law models, with
$\chi^2/n = 344/213$; it has a window energy (source-frame) of $0.78
\pm 0.01$ keV and an absorbing column of $(3.7 \pm 0.2) \times
10^{21}$ cm$^{-2}$, and the best-fitting power-law index is very flat
($\alpha = 0.10 \pm 0.02$). The ionized absorber model gives a
slightly better fit, $\chi^2 = 322/213$, with the ionizing continuum
power-law index fixed to $\alpha_i = 0.5$, the absorber temperature
set to $3 \times 10^4$ K and the iron abundance set to unity. In this
model, shown in Fig.\ \ref{351core}, the power-law index $\alpha =
0.42 \pm 0.04$, the absorbing $N_{\rm H} = (1.5 \pm 0.1) \times
10^{22}$ cm$^{-2}$, and the ionization parameter $\xi = 34 \pm 5$ ergs
cm$^{-1}$. This absorbing column is consistent with the value
obtained (under a slightly different model) by \citet{nfpe99}.

None of these models is a particularly good fit to the data, with
particularly strong scatter about the models to be seen below 0.6 keV (Fig.\
\ref{351core}). In Table\ \ref{restab} we tabulate the results for the
two-power-law models, as they remain formally the best fits. Two weak
line-like features are seen in the spectrum, at observed energies of
about 3.2 and 4.7 keV (Fig.\ \ref{351core}). When these are fit with
Gaussian models, the first feature's width is not well constrained,
and it is best fitted with very broad lines. This reduces our
confidence in its physical reality as an emission feature. The second
feature can be modeled as a Gaussian with rest-frame energy $6.40 \pm
0.07$ keV and $\sigma = 70 \pm 70$ eV, with equivalent width 70 eV.
The improvement to the fit of adding this feature is very limited, but
given its energy it may be a marginal detection of the iron K$\alpha$
line.

\subsection{Hotspots}

\begin{figure*}
\plotone{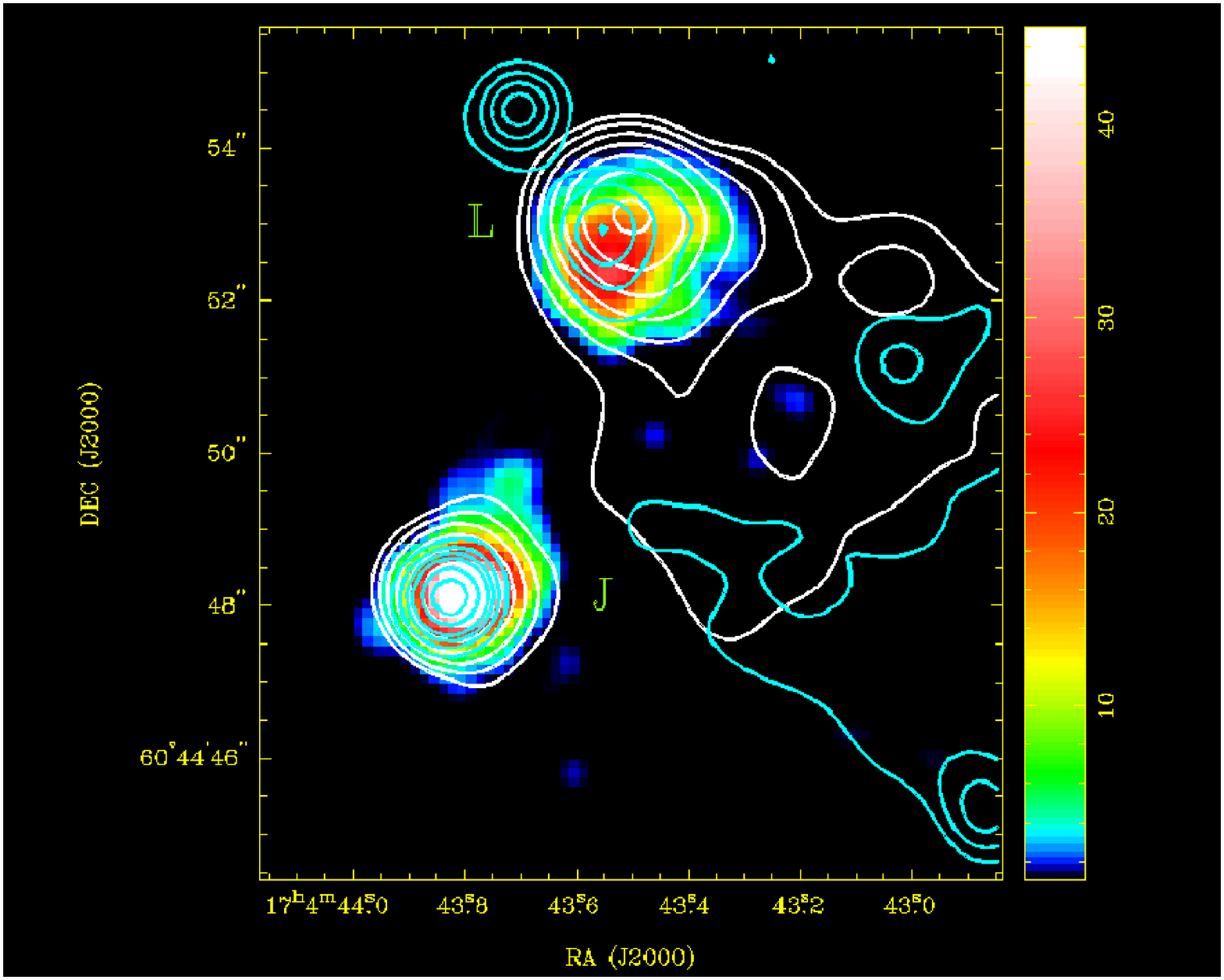}
\caption{3C\,351 X-ray hotspots. The {\it Chandra} data are smoothed
with an $0\farcs5$ Gaussian. The white contours are 8.4-GHz radio
contours with a resolution of $0\farcs85$, approximately matched to
the X-ray resolution. The contours increase logarithmically by a
factor 2 and the lowest contour is at 0.25 mJy beam$^{-1}$. The light
blue contours are the {\it HST} $R$-band data at the same resolution,
and also increase logarithmically by a factor 2 (arbitrary units). The
structure in the bottom right of the image is due to noise on the {\it
HST} PC chip.}
\label{351hotspots}
\end{figure*}

As previously reported by \citet{bbcp01}, both of the northern
hotspots of 3C\,351 are detected in the X-ray (Fig.\
\ref{351hotspots}); the compact `primary' hotspot J and the diffuse
`secondary' hotspot L (using the notation of B94). In our observations
five times more counts are obtained than in the dataset used by
\citeauthor{bbcp01}, giving us
well-constrained hotspot spectra (Table \ \ref{restab}). We obtain
1-keV flux densities and spectral indices that agree with the values
determined by \citeauthor{bbcp01} within the joint $1\sigma$
uncertainties.

When these hotspots are compared with a radio map (Fig.\
\ref{351hotspots}) there is a clear offset of about 1\arcsec\ between
the X-ray and radio peaks in the secondary hotspot L, in the sense
that the distance between the peaks of J and L is smaller in the X-ray
than in the radio. L appears to be resolved by {\it Chandra} and to be
comparable in size to its radio counterpart ($\sim 2''$). There is no
clear colour gradient in the X-ray images of L, and no obvious radio
spectral index gradient at high frequency. There is an extension of J
to the north with no radio counterpart, but it is otherwise not
resolved by {\it Chandra}.

\begin{figure*}
\plotone{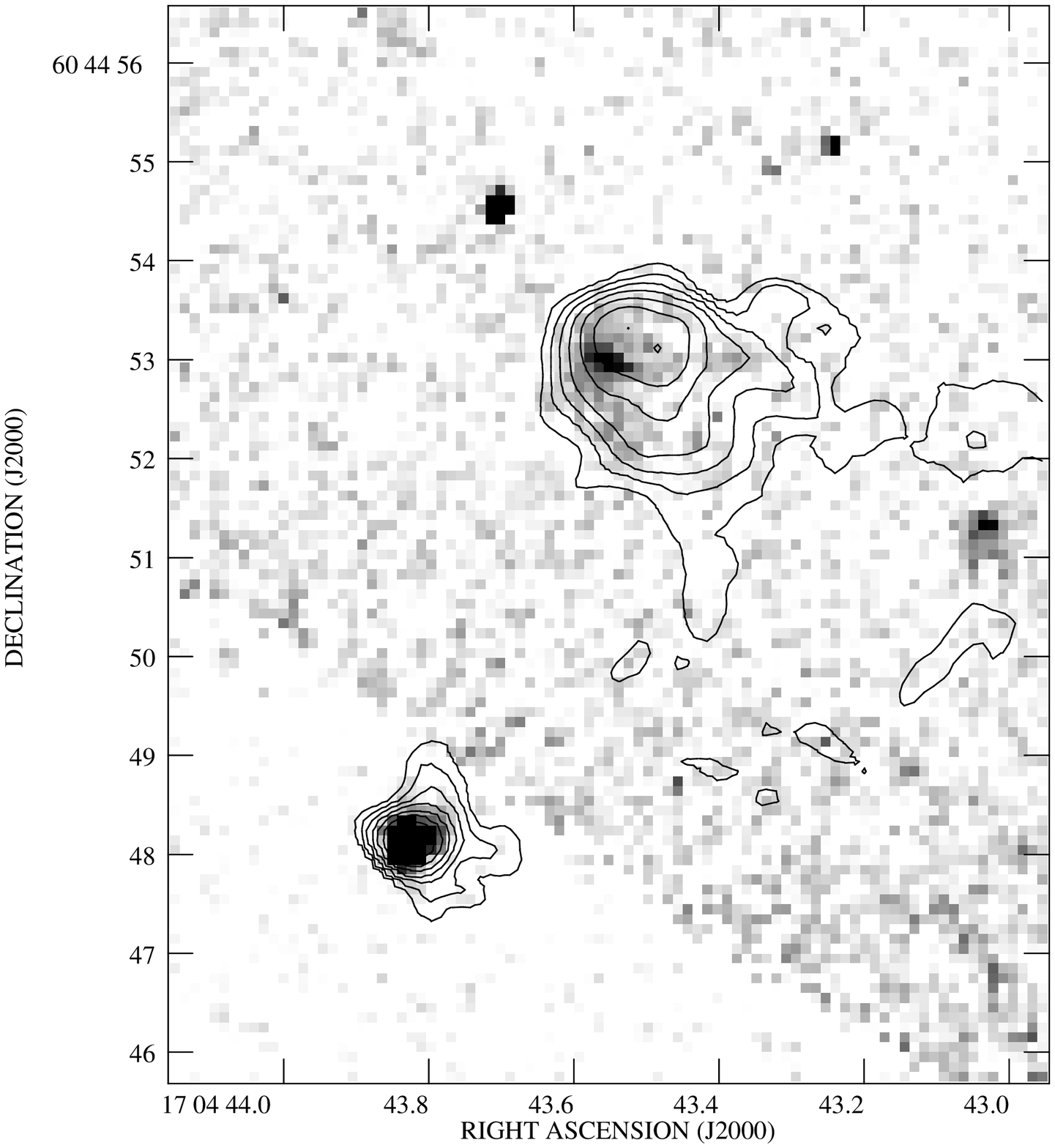}
\caption{3C\,351 optical hotspots at the full {\it HST}
resolution. The {\it HST} data are shifted so that the positions of
hotspot J align in the optical and radio. The contours are from an
8.4-GHz radio map at $0\farcs34 \times 0\farcs23$ resolution. The WFC chip
lies in the bottom left of this image, the PC chip in the top right.}
\label{351hst}
\end{figure*}

Optical counterparts to both northern hotspots in 3C\,351 were
discovered by \citet{r89} and their relationship to the radio emission
confirmed by \citet{lv99}. Further optical
observations were made by \citet{bbcp01}, who also discuss the
{\it HST} observations that we use in this paper. When the {\it HST}
observations are compared with high-resolution radio maps, it appears
that the optical counterpart of hotspot L may be offset from the radio
peak in the same sense as is seen in the X-ray (Figs \ref{351hotspots}
and \ref{351hst}).
The two optical hotspots unfortunately lie on different chips of the
{\it HST} WFPC2, and there are no reference sources in the PC chip to
tie the radio and optical frames together. Significant offsets are
present using the default {\it HST} astrometry, as expected. The
evidence for an offset relies on the accuracy of the {\sc wmosaic}
task in the STSDAS package in {\sc iraf}, used to produce the mosaiced
image shown in these Figures. However, the size of the offset
($0\farcs5$) seems too large to be accounted for by uncertainties in
the WFPC2 chip geometry; other sources which span chip boundaries in
this field (notably some faint galaxies near the quasar nucleus) do so
without showing any evidence for such offsets. We conclude that the
observed offset may well be real. In this case, the optical peak of L lies
between the radio and X-ray peaks.

We agree, within the errors, with the {\it HST} R-band
flux density measured by \citet{bbcp01} for hotspot
J. However, we obtain a substantially lower flux density for hotspot
L, although the errors are comparatively large because of
uncertainties in background subtraction. We adopt flux densities of
$2.4 \pm 0.1$ $\mu$Jy and $1.9 \pm 0.2$ $\mu$Jy for J and L respectively.

\begin{figure*}
\plottwo{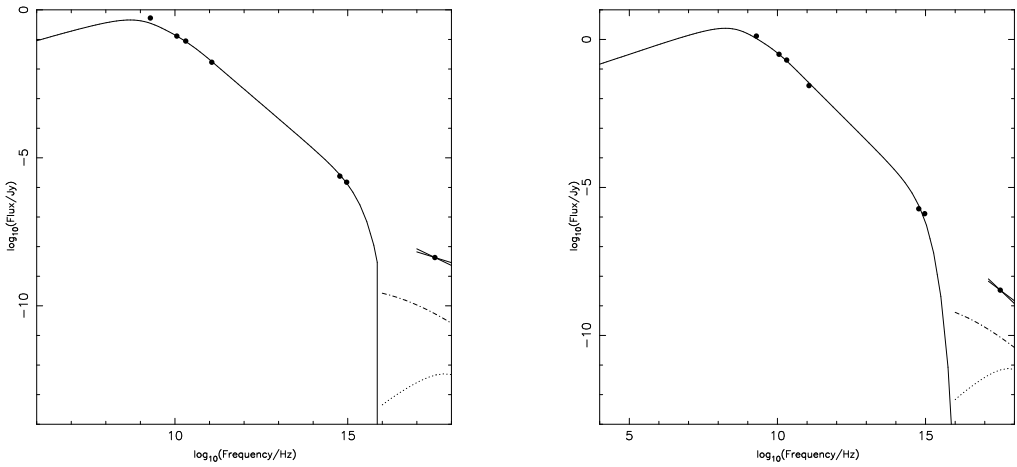}{f16b.eps}
\caption{The source-frame spectra of the N hotspots of 3C\,351: left
is hotspot J and right is hotspot L. Radio points from Table
\ref{radiofluxes}, the R-band {\it HST} flux densities determined by
us, the ground-based B-band flux densities of \citet{bbcp01}
and the X-ray flux densities and spectra are plotted together with the
synchrotron models discussed in the text (solid line) and the
predicted SSC (dot-dashed line) and CMB/IC (dotted line) emission from the
hotspot at equipartition. The high-energy cutoff is set to
$\gamma_{\rm max} = 1.3 \times 10^6$ in this figure.}
\label{351-sed}
\end{figure*}

The compact hotspot of 3C\,351 (J) is not well resolved at any radio
frequency available to us, and is compact even in the {\it HST}
images. The size derived from fitting a homogeneous sphere model
depends on the frequency and resolution of the data used, suggesting
that there may be unresolved spatial or spectral structure. The value
we adopt ($r = 0\farcs16$) is based on fitting a homogeneous sphere
model to the 8.4-GHz data of G02. This is larger than the value used
by \citet{bbcp01}, who just took half the FWHM of the Gaussian
fit of B94. Spectrally, the hotspot seems to have a steeper spectrum
between 1.4 and 8 GHz than between 8 and 15 GHz, but this may be
due to contamination from more extended emission in the low-resolution
1.4-GHz map; the maps of B94 and G02 both show structure around this
hotspot. The high-frequency (15-85 GHz) spectral index of the hotspot
is close to 1.0. If we assume that some of the 1.4-GHz flux density is
not from the compact region, then the 8.4, 15 and 85-GHz and optical
data points can be fit, though not particularly well, with a version
of our standard spectral model, in which the synchrotron spectral
index steepens from 0.5 to 1.0 in the radio, and then retains this
value out to beyond the B-band. Since the optical to X-ray spectral
index is close to 1.0, the 1-keV X-ray data point alone can then be
fit as an extension of the synchrotron spectrum, with $\gamma_{\rm
max} > 6 \times 10^7$ at equipartition (this conclusion differs from
that of \citeauthor{bbcp01}, who find a somewhat lower maximum
contribution from synchrotron emission, because of the steeper
injection index they use). However, the flat X-ray spectrum of the
hotspot ($\alpha_{\rm X} = 0.5 \pm 0.1$) is inconsistent with a synchrotron
model. Like \citeauthor{bbcp01}, we prefer a synchrotron model in which
the break is at higher frequencies and the synchrotron emission is
cutting off in the optical ($\gamma_{\rm break} = 5800$, $\gamma_{\rm
max} = 1.3 \times 10^6$ at equipartition). This improves the fit to
the radio and optical data. However, in this model, plotted in Fig.\
\ref{351-sed}, the predicted inverse-Compton flux density at
equipartition ($B_{\rm eq}$ = 21 nT) is 55 pJy, almost two orders of
magnitude below the observed value. The magnetic field must be reduced
by a factor $\sim 12$, to 1.7 nT, to produce the observed emission
from the SSC process. In this case $\gamma_{\rm max} = 4.8 \times
10^6$, $\gamma_{\rm break} = 2.0 \times 10^4$, and the predicted X-ray
spectral index at 1 keV is 0.6, close to the observed value.

Hotspot L is well resolved in the radio and {\it HST} images. A direct
measurement on the radio map shown in Fig.\ \ref{351hotspots} gives
dimensions for the bright region of about $2\farcs2 \times 1\farcs6$.
No simple geometrical model is a good fit to the structure of this
hotspot, with its off-center brightness peak and filamentary
extensions to the SW. We begin by treating the brightest part of the
hotspot as a homogeneous sphere with $r=0\farcs8$; this radius agrees
both with the high-resolution measurements and with fits to the
low-resolution 1.4 and 15-GHz maps, but is again somewhat larger than
that used by \citeauthor{bbcp01}. The radio and optical data can then be
fit with a spectral model very similar to that used for hotspot J; the
1.4-GHz data point again lies above the extrapolation of the spectrum
inferred from the high-frequency radio data, but neglecting this, we
find very similar break and cutoff values, $\gamma_{\rm break} =
8400$, $\gamma_{\rm max} = 1.3 \times 10^6$, at equipartition ($B_{\rm
eq} = 7.1$ nT). This model is plotted in Fig.\ \ref{351-sed}. If the
X-ray emission is SSC, the predicted inverse-Compton flux density in
this model is 80 pJy, and the magnetic field strength would have to be
reduced by a factor $\sim 9$, to 0.8 nT, to produce all the observed
emission by the inverse-Compton process. The predicted 1-keV X-ray
spectral index in this model is also 0.6, which is somewhat flatter
than the observed value of $0.85 \pm 0.1$.

The inverse-Compton model cannot explain the observed offsets between
the X-ray and radio centroids of hotspot L. To explain this offset
while retaining a simple model of the electron distribution we would
need an external illuminating source, but no such source is apparent; in
particular, hotspot J is much too far away to produce a significant
effect (the predicted flux density from IC scattering of J's photons
by L is only 2 pJy at equipartition, or 2\% of the predicted SSC flux
density). Otherwise, if the X-rays are to be SSC emission, there must
be significant electron spectral structure in the hotspot which we
have failed to take into account in our model. With only one
high-resolution, high-frequency radio map, it is impossible to test
this suggestion, but the possible offset seen between the radio and
{\it optical} positions may support it.

Unlike \citeauthor{bbcp01}, because of the lower R-band optical flux
density we obtain, we find 3C\,351 L to have a relatively flat optical
spectral index, $\alpha_{\rm O} \approx 0.8$ (similar to the X-ray spectral
index), and this means that a synchrotron model connecting the optical
and X-ray emission cannot be ruled out in this hotspot either, and is
an alternative explanation for the bright X-ray emission, though it
would require a second, flat-spectrum synchrotron component to be
present.

The differences we find between the equipartition/minimum energy and
SSC fields in these two hotspots are larger than the factor $\sim 3$
inferred by \citet{bbcp01} because of differences in our assumptions.
As noted above, we use larger angular sizes for the hotspots, and this
accounts for a substantial part of the difference. \citeauthor{bbcp01}
used a different definition of the minimum energy, obtaining a lower
field strength than our calculation would have given on the same
assumptions about hotspot size. They also used a version of the more
complex electron energy spectrum described by \citet{bbcs02a}. This
illustrates a general point: differences in model parameters can have
an important effect on the derived magnetic field strength.

The weak S radio hotspot of 3C\,351 is not detected in the X-ray.
Based on the background count rate around the hotspot region, we can
set an upper limit on its X-ray flux density (Table \ref{restab}). It
does not lie in any of the {\it HST} fields, so no optical upper limit
is available, and we do not have enough high-resolution data to
comment on its radio spectrum. In the 8-GHz maps it is resolved into
two components, with flux densities of 2.2 mJy and 1.1 mJy. Both
components are extended. Using a spectral model similar to that used
for hotspot J, we find that the two components of A taken together
would have an SSC flux density of approximately 0.2 pJy, a factor 250
below our upper limit. Thus, even if these hotspots were brighter than
their equipartition flux density by the same factor as hotspot J, we
would not have detected them in our observations. We infer the weak
constraint $B \ga 0.04 B_{\rm eq}$.

\subsection{Lobes}

\begin{figure*}
\plotone{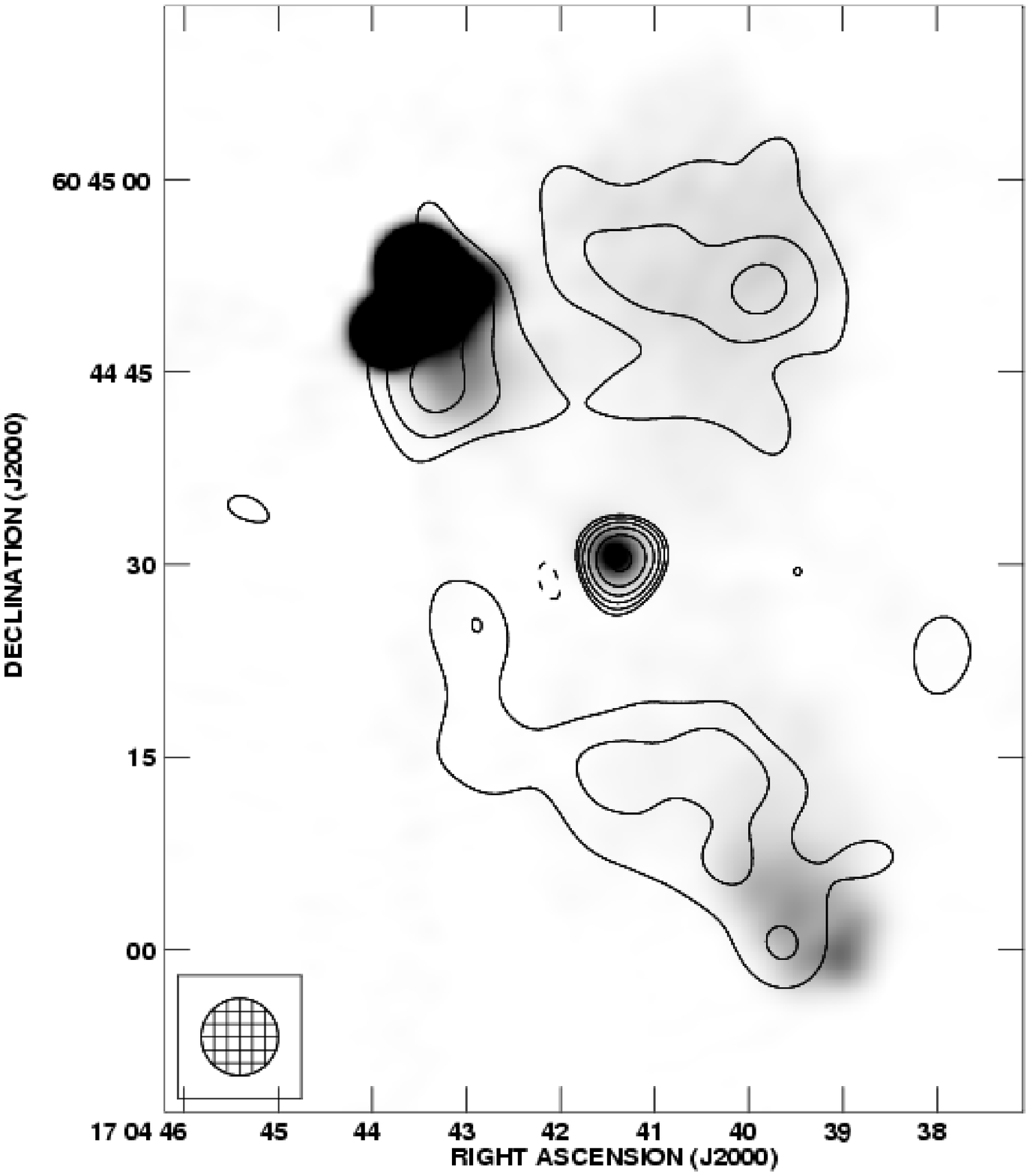}
\caption{The lobes of 3C\,351. The contours show the X-ray emission
smoothed with a 6\arcsec\ (FWHM) Gaussian. The hotspot regions and
background X-ray point sources were masked out before smoothing (some
extended emission is present around the hotspots and gives rise to the
contours near them) and a scaled, smoothed, energy-weighted PSF model
was subtracted from the image after smoothing to remove the nuclear
emission. A narrow peak at the core position shows that the PSF model
is slighly broader than the real data. The lowest contour is the
$3\sigma$ level, defined according to the prescription of \citet{h00}, and the contours increase logarithmically by a factor
$\sqrt{2}$. Negative contours (illustrating the residuals due to core
subtraction) are dashed. The FWHM of the convolving Gaussian is shown
in the bottom left of the image. The greyscale shows the 8.4-GHz VLA
image with 3\arcsec\ resolution of G02; black is 10 mJy beam$^{-1}$.}
\label{351lobe}
\end{figure*}

Both lobes of 3C\,351 are detected with {\it Chandra} (Fig.\
\ref{351lobe}). To extract spectra, we used two circular regions at
identical distances (25\arcsec ) from the nucleus, centered on the
lobe emission, and a matched off-source background region in a
suitable position angle. There are 60 net counts in the northern lobe
and 40 in the southern lobe; this is insufficient for spectral
analysis of either lobe alone but allowed us to fit a model to the
combined emission from both (Table \ref{restab}).

Modeling the lobes as described above for 3C\,263 and 3C\,330, we find
a predicted 1-keV flux density for the CMB/IC process of 0.21
nJy for the N lobe and 0.17 nJy for the S lobe, respectively factors
$\sim 5$ and $\sim 4$ below what is observed. If the X-ray emission
from these lobes is to be produced by the CMB/IC process, the magnetic
field strength must be a factor 2--2.5 below equipartition. The
pressures in the lobes are then $\sim 4 \times 10^{-13}$ Pa.

\subsection{Extended emission}

There is no evidence for extended, cluster-related X-ray emission in
3C\,351. No additional extended model that we have fitted improves the
fitting statistic significantly, and the best-fitting models contain
only a few tens of counts. If we take these as very rough upper
limits, the X-ray luminosity of 3C\,351's environment is no more than
a few $\times 10^{42}$ ergs s$^{-1}$ (depending on choice of
temperature and abundance), a group environment at best. Other
evidence, such as galaxy count studies \citep[e.g.,][]{hesr01},
suggests that 3C\,351's environment is poor, so this is not a
surprising conclusion. The gas in groups of galaxies of this sort of
luminosity can have a pressure at 100-kpc radii comparable to that
estimated above for the lobes of 3C\,351, so it remains possible that
3C\,351's lobes are pressure-confined by a faint X-ray emitting
environment.

\section{Discussion}

\subsection{Evidence for and against equipartition in hotspots}
\label{hsd}

\begin{deluxetable}{lrrrr}
\tablewidth{10cm}
\tablecaption{The ratio between equipartition and inferred field
strengths in hotspots and lobes}
\tablehead{Source&\multicolumn{4}{c}{$B_{\rm eq}/B$}\\
&N hotspot&S hotspot&N lobe&S lobe}
\startdata
3C\,263&$<4$&$1.9 \pm 0.1$&$2.0 \pm 0.3$&$2.0 \pm 0.4$\\
3C\,330&$1.0 \pm 0.1$&$1.5 \pm 0.4$&$1.4 \pm 0.3$&$1.4 \pm 0.3$\\
3C\,351&$12 \pm 0.05$&$<25$&$2.5 \pm 0.4$&$2.0 \pm 0.5$\\
\enddata
\label{lobehot}
\tablecomments{For 3C\,351, hotspot J is used. Errors are based on the
statistical errors on 1-keV flux density quoted in Table \ref{restab},
and do not include systematic uncertainties in the modelling.}
\end{deluxetable}

The inferred field strengths, relative to the equipartition values, of the
hotspots of the target sources are listed in Table \ref{lobehot}. Of
the three sources, one (3C\,330) has hotspots whose X-ray emission is
consistent with SSC emission at equipartition, one (3C\,263) has an
inferred hotspot magnetic field strength a factor 2 below the
equipartition value (consistent with other sources, such as 3C\,295,
which show a similar deviation from equipartition) and one (3C\,351)
has hotspots whose X-ray emission, if it were of SSC origin, would
imply a magnetic field strength an order of magnitude below the
equipartition value.

What process is responsible for the anomalously bright, flat-spectrum
hotspots in 3C\,351? We can consider several classes of explanation.
The first involves retaining the SSC process, with some modifications;
the second involves invoking relativistic beaming, an idea supported
by the observation of very bright, one-sided radio hotspots on the jet
side of 3C\,351 and by the displacement of the hotspots with respect
to the radio lobe; and the third involves some emission mechanism
other than the inverse-Compton process. We discuss several variants of
these models in turn.

\begin{itemize}
\item[1a] {\it SSC with standard electron spectral model, $B \ll B_{\rm eq}$}.
As discussed above, this model does not explain the offsets between
the radio and X-ray peaks in hotspot L.
\item[1b] {\it SSC with standard electron spectral model, $B \ll
B_{\rm eq}$, and spatial variation of the electron population}. This
model can explain the offsets between the peaks in various wavebands,
at the cost of introducing new features in the electron population for
which there is no independent observational evidence.
\item[1c] {\it SSC with second electron population, $B \sim B_{\rm
eq}$}. If we introduce an additional population of low-energy
electrons whose synchrotron emission is below the observed radio band,
, we can greatly increase the SSC flux. If we also allow this population to be offset
with respect to the radio-emitting population, we can explain the
properties of hotspot L. This model has the same disadvantages as 1b.
\item[1d] {\it SSC with relativistic beaming, $B \sim B_{\rm eq}$}.
Beaming would mean that our estimate of the hotspot flux density would
not correspond to the source-frame value. The predicted SSC flux
density and the ratio $B_{\rm eq}/B$ depend on the Doppler factor
$\cal D$ (${\cal D} = [\Gamma(1 - \beta \cos \theta)]^{-1}$ for a
hotspot, where $\theta$ is the angle to the line of sight and $\Gamma
= (1-\beta^2)^{-1/2}$ is the bulk Lorentz factor). Beaming causes the
observed flux densities to change by a factor ${\cal D}^{(3+\alpha)}$.
For a pure power-law spectrum with electron energy index $p$ ($\alpha
= (p-1)/2$), neglecting aberration effects, the predicted SSC flux can
be shown to go as ${\cal D}^{-12/(p+5)}$, and so, for values of $p$ in
the range 2--3, Doppler boosting, (${\cal D}>1$) reduces the predicted
equipartition flux while Doppler dimming (${\cal D}<1$) increases it.
To place 3C\,351 at equipartition we would require $\beta \ga 0.999$
($\Gamma \ga 20$) with the source lying close to the plane of the sky.
This is clearly not a plausible model for a quasar; in addition, the
rest-frame radio luminosity of the hotspot becomes very large.

\item[2a] {\it Boosted CMB, $B \sim B_{\rm eq}$}. Relativistic fluid speeds
in or close to the hotspot would give rise to an increased contribution from
inverse-Compton scattering of the microwave background, which is
negligible in all our sources if relativistic effects are not present.
The effective energy density in microwave background photons increases
as $\sim \Gamma^2$ \citep[e.g.,][]{ds93}, but this is partly
offset by the reduction in the electron number density. There are
additional corrections because of the anisotropic nature of the
inverse-Compton process; the net result is that a source exhibiting
emission from this process must be close to the line of sight. Because
of the blueshifting of the CMB photons in the source frame, the
results are also strongly dependent on the presence of low-energy
electrons; in order to obtain a spectrum which is not flat or
inverted, we require, approximately, $2 \times 10^{-7}
(1+z)\gamma^2_{\min} \Gamma^2 \ll 1$. To reproduce the observed
emission from 3C\,351 with an equipartition $B$-field and the lowest
possible bulk $\Gamma$, $\Gamma \sim 5$, we require $\gamma_{\rm min}
\sim 10$ and $\theta = 0\degr$. Larger angles to the line of sight
require larger $\Gamma$ values. This model thus requires 3C\,351 to be
very close to the line of sight (for $5 < \Gamma < 20$, $\theta \la
6\degr$, so the linear size of the source would have to be $\ga 4$
Mpc, if bending is neglected). We also require highly relativistic
flows to power both J and L: either the jet splits, or (more
conventionally) a relativistic outflow from J powers L. In the second
case, we would expect $\Gamma$ to be different in the two hotspots.
The angles between the line of sight and the velocity vectors of the flows powering the hotspots are likely to differ in either situation,
because the hotspots and the core do not lie on a straight line: so in
this model the parameters must `conspire' to ensure that the X-ray to
radio ratios remain similar in the two hotspots. This model does not
explain the offsets in hotspot L without some additional assumptions,
such as finely tuned velocity vector variations in the hotspot.
\item [2b] {\it Boosted CMB, but hotspots are `jet knots', $B \sim
B_{\rm eq}$}. In this model, the objects we have described as hotspots
are not terminal hotspots at all, but knots in a projected jet (which
presumably terminates somewhere else (e.g., in the displaced N lobe)
in a hotspot of brightness comparable to that in the S lobe. This
model has the advantage that we do not (necessarily) expect jet
deceleration between knots, and that it allows us to retain
sub-relativistic speeds in the hotspots. Otherwise it has the same
disadvantages as 2a.
\item [2c] {\it Boosted hotspot back-scattering, $B$ unknown}. We cannot
rule out the possibility that the X-ray emission does not originate in
the hotspot (post-shock) region at all. It could instead be due to
scattering of hotspot photons by the incoming, presumably highly
relativistic jet. This model has the advantage that it can explain the
offsets observed in hotspot L. It has the disadvantage that, as no
radio jets are observed entering the hotspot, we cannot use the
observations to constrain magnetic fields or jet speeds. Because the
IC emission from this process is even more anisotropic than for 2a and
2b, the problem of the angle made by the jet to the line of sight
becomes greater in this model.
\item[3a] {\it Synchrotron from second electron population, $B$
unknown}. A second, flat-spectrum synchrotron component in the X-ray
(and possibly also the optical) can explain many of the observations
of 3C\,351's hotspots, including the offsets seen in hotspot L if the
second electron population is spatially offset from the low-energy
electrons responsible for the radio emission. Given the very different
loss timescales in the radio and X-ray, and the rapidly changing
nature of hotspots (and the bulk flows that feed them) that numerical
simulations suggest, there is no very strong reason to believe that a
single electron energy power law {\it should} describe observations
made at a given moment, so a picture of this kind is not inconsistent
with standard models of radio sources. The X-ray emission in the jets
of certain sources, such as 3C\,120 \citep{hhss99} and 3C\,273
\citep{jrmp02}, has been explained in terms of a model of this kind.
\item[3b] {\it Synchrotron from second electron population and SSC, $B$ unknown but $B \sim B_{\rm eq}$
permitted}. This is a trivial variation on model 3a in which some of
the X-ray emission is due to SSC rather than synchrotron.
\item[3c] {\it Proton-induced cascade or other exotica, $B$ unknown}. As
usual, we cannot rule out the possibility that some less familiar
emission process makes a contribution to the X-ray emission.
\end{itemize}

All these models are {\it ad hoc} in one way or another. The beaming
models do help to explain the large-scale asymmetry of the radio
source as well as the X-ray emission, but they all require 3C\,351 to
be very close to the line of sight and extremely large in unprojected
linear size, as well as invoking speeds in the hotspot regions that
are very much higher than have been inferred from other work.
The other models require less extreme parameters for the radio source,
but do not answer the question of why 3C\,351 is unlike the other
sources in the sample, which show no evidence for strong departures
from equipartition, extra electron populations, or new emission
mechanisms.

It is interesting to compare the results on 3C\,351 with those on
other FRII sources with `bright' X-ray hotspots, such as Pictor A and
3C\,390.3. In the case of Pictor A, \citet*{wys01} found that a
magnetic field a factor 14 below the equipartition value would
reproduce the flux density of the X-ray hotspot, but they reject a
simple SSC model of the type we have used in this paper because of the
well-constrained steep spectrum of the hotspot ($\alpha_{\rm X} = 1.07 \pm
0.11$). A simple synchrotron model is also rejected because the
optical data require a cutoff in the optical \citep*{myr97}. However,
\citeauthor{wys01} do not rule out a model in which there are
contributions from both X-ray synchrotron and SSC emission, as in
model 3b above, which requires $B_{eq}/B \approx 9$. This factor is
similar to those that we would require in 3C\,351 on the pure SSC
model (1a). They also consider the possibility of a second electron
population (models 1c and 3a above).

In the case of 3C\,390.3, \citet{hll98} rejected a
SSC model for the {\it ROSAT}-detected X-ray hotspot, hotspot B,
because the X-ray emission was much brighter than their SSC
prediction. The {\it Chandra} observation of this source
(Wagner \etal , in preparation) detects all three compact hotspots
in 3C\,390.3 and allows a more precise determination of some of the
source parameters.  Examining the archive data, we find that the
brightest hotspot, hotspot B, has a 1-keV flux density of $4.5 \pm
0.3$ nJy, in good agreement with the result of \citeauthor{hll98}, and
$\alpha_{\rm X} = 0.88 \pm 0.17$. The radio, optical and X-ray data in hotspot
B can be connected with a simple synchrotron model. An SSC model
requires a large departure from equipartition; with our cosmology and
standard spectral assumptions we find $B_{\rm eq}/B \approx 90$, much
higher than in any other source we have examined. Moreover, the
predicted SSC spectral index is $\sim 0.5$, while the spectral index
in the synchrotron model is $\sim 1.0$, so that the {\it Chandra}
spectrum favors the synchrotron model. We see no reason to believe
that there is a significant contribution from SSC emission in hotspot
B of 3C\,390.3. The {\it Chandra} spectra of the two
other X-ray detected hotspots in 3C\,390.3 are even steeper than in
hotspot B, which suggests that a synchrotron model is required there
as well. It seems likely that the differences between the radio to
X-ray spectra of the three hotspots, remarked on by \citeauthor{hll98}, can
be explained in terms of similar electron spectra with different
high-energy cutoffs. Given the large number of jets in low-power radio
galaxies now thought to be due to synchrotron radiation (\S 1), the
existence of the required high-energy electrons in hotspots is no
longer as remarkable as it once seemed. This observational evidence for
synchrotron emission in hotspots is indirect support for one of the
synchrotron models (3a/b above) to explain the emission from 3C\,351.

The {\it Chandra} data on hotspots in general are therefore
reinforcing the conclusion, already suggested by {\it ROSAT}
observations, that more than one emission mechanism is operating in
the X-ray in FRII hotspots. Some hotspots, of which 3C\,390.3 is to
date the most convincing example, require an X-ray synchrotron model.
Some, such as Cygnus A, 3C\,123, 3C\,263 and 3C\,330, are likely to be
pure SSC, with no significant contribution to the X-rays from
synchrotron emission and a hotspot magnetic field close to the
equipartition value. And some, such as Pictor A and perhaps 3C\,351
(model 3b above), may have synchrotron and SSC emission at similar
levels, or may be different from the other sources in some more
interesting way. In order to distinguish between these cases, good
radio and optical information and, ideally, an X-ray spectrum of the
hotspot are required: further high-resolution observations in the
radio, optical and UV may be required to to disentangle the emission
mechanisms in Pictor A and 3C\,351.

In this picture, the questions remaining to be answered are:
\begin{itemize}
\item What is special about the hotspots that exhibit X-ray
synchrotron emission?  It's particularly striking that all three
compact hotspots of 3C\,390.3 are X-ray synchrotron emitters, despite
being quite different in their local environments and radio structures
\citep[see][]{lp95}. There is no obvious way in which these hotspots,
in quite different parts of the source, can `know' that they should
accelerate electrons to high energies --- unless some intrinsic
property of the beam is involved.
\item What is special about the hotspots, like 3C\,351 and possibly Pictor A,
that may have $B_{\rm eq}/B \gg 1$?
\end{itemize}

\citet{bbcp01} suggested a partial answer to the second of
these questions. They pointed out that, in standard shock-acceleration
models, the high-energy cutoff of the synchrotron electron population
is higher for lower hotspot magnetic fields, because it is determined
by the balance between acceleration and synchrotron loss processes. In
other words, we might expect to see an enhanced contribution from SSC
emission in radio sources which we know by other means to have a
synchrotron spectrum extending to high energies. This might apply to
3C\,351 and Pictor A, which are known to have optical synchrotron
hotspots, while sources like Cygnus A and 3C\,123, whose electron
spectra are known to cut off at lower energies, have fields closer to
equipartition. However, one source in the current sample,
3C\,263, breaks the trend by having an optical synchrotron hotspot
without being particularly X-ray bright. And this picture does not
explain {\it why} it is that some sources have particularly low
hotspot B-fields, or why, in the case of 3C\,351, {\it both} hotspots
appear to do so.

Numerical simulations \citep*[e.g.,][]{tjr01} lead us to expect
that hotspots will be transient structures, and may well have
different magnetic field and electron acceleration behaviour at
different times. All hotspots may pass through phases of
inverse-Compton domination and synchrotron domination of their X-ray
output. If this is the case, neglecting the probably important
question of observational bias, then the statistics of detected
sources are telling us that sources spend approximately equal times in
the two regimes. However, this does not explain (and is even to some
extent incompatible with) the observation of correlated properties in
spatially separated hotspots, as observed in Cygnus A, 3C\,351 and
3C\,390.3.

\subsection{Inverse-Compton emission from lobes? Constraints on lobe physics}
\label{lobesect}

Both lobes of all three sources are detected in the X-ray. In all
three cases, the measured X-ray emission is close to the predicted
value for inverse-Compton scattering of CMB photons, requiring field
strengths which are factors between 1.4 and 2.5 below equipartition
values on simple models which treat the lobes as uniform (Table
\ref{lobehot}). We have not taken into account any possible
illumination of the lobes by the quasar nuclei (nuclear
inverse-Compton, NIC), which would give an additional photon
contribution and reduce the discrepancy between the inferred and
equipartition magnetic field strengths. We do not believe that nuclear
illumination is likely to dominate the CMB photon field in our
sources, since in 3C\,263 and 3C\,351 there is no evidence for a
higher ratio of X-ray to radio flux in the lobes on the counterjet
side, as would be expected, due to the anisotropy of the
inverse-Compton process, in a NIC model \citep{bsc97}.

Is the lobe emission inverse-Compton in origin at all? The similarity
of the observed fluxes to the predicted values, and the tight
relationship between the emitting regions in radio and X-ray, suggest
this, but cannot confirm it. The X-ray spectra are not good enough in
any of our sources to allow us to rule out a thermal origin for the
lobe emission, although the existence of some hard counts (e.g., in
3C\,263's N lobe and 3C\,351) would imply quite high temperatures, $kT
\sim 5$ keV). If the X-rays were thermal, their spatial similarity to
the radio emission means that they would have to originate either
inside the lobes, or in a boundary layer around the lobes with a
thickness considerably less than the lobe dimensions (i.e., no more
than a few kpc, given the spatial resolution of {\it Chandra}).
Polarization observations are conventionally taken to rule out the
presence of large amounts of thermal material inside radio lobes,
although, as pointed out by \cite{l84}, large amounts of thermal
material can be hidden by suitable magnetic field configurations.
Unfortunately, estimates of the depolarization measure $DP$ are not
available to us for either of our quasars. For 3C\,330, \cite{f01}
quotes values of $DP$ (the ratio of polarization at two frequencies,
in this case 1.4 and 5 GHz) of 0.84 and 0.55 for the two lobes. (A
lower $DP$ means more depolarization.) We can convert the observed
{\it Chandra} count rate for the lobes of 3C\,330 into a thermal
proton density using {\sc xspec} and assuming that the thermal plasma
uniformly fills the radio lobes; we use a {\sc mekal}
model\footnote{This model
includes both thermal bremsstrahlung and line emission from a hot
optically thin plasma, and is generally found to be a good fit to high
signal-to-noise observations of clusters.} with $kT =
5$ keV, $0.3$ solar abundance and Galactic absorption. We find that a
thermal proton density of $\sim 10^{4}$ m$^{-3}$ would be required.
Using equipartition fields, this would give rise to a large Faraday
depth, and, using the results of \citet{cj80}, we would
expect $DP \sim 0.08$, which is much less than the observed value; we
would also expect a very low degree of polarization at low
frequencies, whereas we know that the source remains polarized at 1.4
GHz. We cannot reduce the Faraday depth by significantly decreasing
the assumed magnetic field strength, as the X-ray observations set a
lower limit on the field strength, $B > B_{\rm eq}/1.5$. So, for
simple field geometries, the idea that thermal protons are inside the
lobes is not consistent with the radio data.

If the thermal protons are instead in a boundary layer of shocked gas
around the source, their density can be estimated in the same way, and
is greater than the value calculated above by the square root of the
ratio of the boundary layer to source volumes: for example, if the
boundary layer were $0\farcs5$ (3.5 kpc) thick in 3C\,330, the density
would be higher by a factor $\sim 1.5$. Here Faraday-rotation
constraints are less helpful, because we do not know the value of the
external magnetic field strength. However, if the external thermal
material is assumed to be shocked ambient gas, well-known results for
the density contrast in a shock imply that the ambient gas density
must be no more than a factor 4 less dense than the inferred external
density, or at least a few $\times 10^{3}$ m$^{-3}$. This is
comparable to, though on average somewhat larger than, the density of
the external material inferred from the observations described in \S
\ref{330ext} at the distance of the lobes of 3C\,330, so this model
remains possible from that perspective. However, it seems likely that
this external gas is too hot (particularly in 3C\,263) to give rise to
the observed X-ray emission when shocked, unless the shocks are only
weak (in which case the stand-off distance needs to be large). There
is also no obvious gradient of X-ray emissivity along the sources,
particularly 3C\,330, although the $\beta$-model fits of
\S\ref{330ext} imply that the external density drops by an order of
magnitude along the length of the lobes in that source. And there is
no evidence in the X-ray data of the limb-brightening expected in such
a model, although the small number of counts from the lobes makes
detailed analysis of the lobe structure difficult. We conclude that,
although models in which the lobe-related X-ray emission are due to an
external boundary layer are not ruled out, they are more contrived
than the alternative of an inverse-Compton origin with a magnetic
field somewhat lower than the equipartition value.

Whether or not the emission is inverse-Compton in origin, its
detection places some interesting constraints on the filling factors
in the lobes, by giving an upper limit on lobe X-ray emissivity. As
pointed out by \citet{hw00}, the dependence of
inverse-Compton emissivity on filling factor is complicated, depending
on the nature of the `fluid' that fills the gaps between
radio-emitting regions. However, if the fluid is relativistic
electrons, i.e., the lobes consist of a more or less uniform
distribution of electrons with a very wide range in magnetic field
strengths, then the inverse-Compton emissivity increases as the
filling factor $\phi$ decreases. In the simple case in which the
magnetic field is either on or off, and where on is in energy
equipartition with the synchrotron-emitting electrons, the IC
emissivity over the entire lobe goes up approximately as
$\phi^{-4/7}$. This sort of model is constrained by the observations
to have $\phi \ga 0.1$. The constraint is difficult to evade if we
require that there are no strong pressure differences in the lobes.
Filling factors lower than this are only possible if the non-radiating
`fluid' does not consist of electrons --- for example, if the magnetic
field dominates the overall energy of the lobes. We note that less
extreme models with inhomogenous magnetic field distributions are not
ruled out by the data. For example, the models used by \cite{t91}
give only a modest increase ($\sim 20$\%) in the expected
inverse-Compton emissivity if the mean field energy is in
equipartition with the electrons.

A CMB inverse-Compton model, and the inferred relatively small but
non-zero departures from equipartition, are consistent with early
observations of less powerful sources with {\it ROSAT} and {\it ASCA}
\citep[e.g.,][]{flkf95, tlsb96, tkmi98} as well as with new results
now emerging from {\it Chandra} observations of sources more similar
to our targets \citep*{bbcs02a,bcdb02b,fcj02}.

\subsection{Thermal emission and pressure balance}

The extended emission around 3C\,263 and 3C\,330 seems very likely to
be thermal emission from a cluster-scale atmosphere: we have thus
confirmed earlier predictions \citep{hegy95} that 3C\,263 should
lie in an X-ray emitting cluster. If the lobe X-rays are indeed
inverse-Compton emission, the model we favor of those discussed in
\S\ref{lobesect} above,
then the required departures from equipartition produce lobe pressures
which are consistent, within the uncertainties on the various
calculations, with the inferred external pressures.  Small
departures from equipartition may therefore solve the problem,
discussed by \citet{hw00}, of the apparent
under-pressuring of the radio lobes with respect to the external
medium; this is particularly true if some of the X-ray emission
detected by \citeauthor{hw00} with {\it ROSAT} was in fact
inverse-Compton emission from the lobes of their target sources. More
observations are required to test this hypothesis. We note, however,
that our observations provide little or no support for the idea that
lobes are strongly overpressured with respect to the external
medium. For this to be the case, there would need to be another
contribution to the internal lobe pressure, for example from
relativistic protons, which cannot be detected with the present
observations.

\section{Conclusions}

We have detected X-ray emission from the hotspots and lobes of three
powerful FRII radio sources, and low-level extended emission from the
hot-gas environments of two of them.

The hotspots in two of the objects we have studied, 3C\,330 and
3C\,263, are consistent with being SSC emission with a hotspot
magnetic field within a factor 2 below the value for energy
equipartition with the synchrotron-emitting electrons. We argue
therefore that magnetic field strengths close to equipartition are
common in hotspots.

The double hotspots in the third source, 3C\,351, are difficult to
accommodate within an SSC model; they require magnetic field strengths
much further from equipartition and, at least in one hotspot, there
are offsets between the radio and X-ray centroids that a simple SSC
model does not explain. The flat X-ray spectral indices rule out the
possibility that the X-rays can be synchrotron emission from the
electron population responsible for the radio emission. We are left
with several alternative models, none of them particularly attractive:
the most plausible condenders are an SSC model with large departures
from equipartition, together with unmodeled spatial structure in the
electron population in the secondary hotspot, a synchrotron model with
a second, high-energy electron population, or strong beaming effects
in the hotspot regions. In any case, it is striking that the two
hotspots behave in such a similar way, in spite of their large spatial
separation (and the expectation that primary and secondary hotspots
should have different physics). As we have pointed out, other X-ray
hotspot sources, like Cygnus A and 3C\,390.3, seem to share this
tendency for the hotspots to appear to know about each other.

The lobe emission we have detected seems most likely to be due to
inverse-Compton scattering of microwave background photons by the
synchrotron-emitting electron population. All three sources can be
explained by such a model, in which the magnetic field strengths in
the lobes are at most a factor 2 below equipartition (on a simple
model with uniform field and electron distributions in the lobes). If
this model is correct, the pressures in lobes are a factor of a few
above the canonical minimum pressures. This factor may be enough to
account for the discrepancy between the minimum pressure and the
pressure in the external hot gas environment, discussed by
\citet{hw00}, and indeed we suggest that the two sources with
detected extended X-ray emission, 3C\,263 and 3C\,330, are in
approximate pressure balance with the hot gas in their cluster/group
environments; such a model is also possible for 3C\,351.

\acknowledgements

We are grateful to Gianfranco Brunetti for helpful discussions of
inverse-Compton scattering, and to an anonymous referee for comments
which improved the presentation of the paper.

The work at SAO was partially supported by NASA grants and contracts
GO1-2109X and NAS8-39073.

The National Radio Astronomy Observatory is a facility of the National
Science Foundation operated under cooperative agreement by Associated
Universities, Inc. We are grateful to the VLA schedulers for allowing
us to make an observation of 3C\,263 at short notice, and to the data
analysts for supporting our archive requests. We thank George Gilbert,
Robert Laing, Laura Mullin and Julia Riley for providing
us with radio maps and $uv$ data at various stages in this project,
and Hugh Aller, Wil van Breugel, Robert Laing and Bev Wills for
allowing us to use their VLA archival observations.


\begin{thebibliography}{}
\bibitem[Arshakian \& Longair(2000)]{al00}Arshakian, T.G., \& Longair, M.S., 2000, \mnras, 311, 846
\bibitem[Best et al.(1995)]{bblr95}Best, P.N., Bailer, D.M., Longair, M.S., \& Riley, J.M., 1995, \mnras, 275, 1171
\bibitem[Birkinshaw \& Worrall(1993)]{bw93}Birkinshaw, M., \& Worrall, D.M., 1993, \apj, 412, 568
\bibitem[Bridle et al.(1994)]{bhlb94}Bridle, A.H., Hough, D.H., Lonsdale, C.J., Burns, J.O., \& Laing, R.A., 1994, \aj, 108, 766
\bibitem[Brunetti(2000)]{b00}Brunetti, G., 2000, Astroparticle Phys., 13, 107
\bibitem[Brunetti et al.(1997)Brunetti, Setti \& Comastri]{bsc97}Brunetti, G., Setti, G., \& Comastri, A., 1997, \aap, 325, 898
\bibitem[Brunetti et al.(2001)]{bbcp01}Brunetti, G., Bondi, M., Comastri, A., Pedani, M., Varano, S., Setti, G., \& Hardcastle, M.J., 2001, \apj, 561, L157
\bibitem[Brunetti et al.(2002a)]{bbcs02a}Brunetti, G., Bondi, M., Comastri, A., \& Setti, G., 2002a, \aap, 381, 795
\bibitem[Brunetti et al.(2002b)]{bcdb02b}Brunetti, G., Comastri, A., Dallacasa, D., Bondi, M., Pedani, M., \& Setti, G., 2002b, to appear in New Visions of the X-ray Universe in the XMM-Newton and Chandra Era, ed. F.\ Jansen et al., ESA Conference series (astro-ph/0202373)
\bibitem[Burbidge(1956)]{b56}Burbidge, G., 1956, \apj, 124, 416
\bibitem[Carilli et al.(1991)]{cpdl91}Carilli, C.L., Perley, R.A., Dreher, J.W., \& Leahy, J.P., 1991, \apj, 383, 554
\bibitem[Cavaliere \& Fusco-Femiano(1978)]{cf78}Cavaliere, A., \& Fusco-Femiano, R., 1978, \aap, 70, 677
\bibitem[Cioffi \& Jones(1980)]{cj80}Cioffi, D.F., \& Jones, T.W., 1980, \aj, 85, 368
\bibitem[Davis(2001)]{d01}Davis, J.E., 2001, \apj, 562, 575
\bibitem[Dennett-Thorpe et al.(1997)]{dbls97}Dennett-Thorpe, J., Bridle, A.H., Laing, R.A., Scheuer, P.A.G., \& Leahy, J.P., 1997, \mnras, 289, 753
\bibitem[Dermer \& Schlickeiser(1993)]{ds93}Dermer, C.D., \& Schlickeiser, R., 1993, \apj, 416, 458
\bibitem[Fabian et al.(2002)Fabian, Celotti \& Johnstone]{fcj02}Fabian, A.C., Celotti, A., \& Johnstone, R.M., 2002, \mnras , submitted (astro-ph/020488)
\bibitem[Feigelson et al.(1995)]{flkf95}Feigelson, E.D., Laurent-Muehleisen, S.A., Kollgaard, R.I., \& Fomalont, E., 1995, \apj, 449, L149
\bibitem[Fernini(2001)]{f01}Fernini, I., 2001, \aj, 122, 83
\bibitem[Fernini et al.(1997)Fernini, Burns \& Perley]{fbp97}Fernini, I., Burns, J.O., \& Perley, R.A., 1997, \aj, 114, 2292
\bibitem[Fiore et al.(1993)]{femw93}Fiore, F., Elvis, M., Mathur, S., Wilkes, B., \& McDowell, J.C., 1993, \apj, 415, 129
\bibitem[Gilbert et al.(2002)]{grpa02}Gilbert, G., Riley, J.M., Pooley, G.G., Alexander, P., \& Hardcastle, M.J., 2002, submitted to MNRAS
\bibitem[Hall et al.(1995)]{hegy95}Hall, P.B., Ellingson, E., Green, R.F., \& Yee, H.K.C., 1995, \aj, 110, 513
\bibitem[Hardcastle(2000)]{h00}Hardcastle, M.J., 2000, \aap, 357, 884
\bibitem[Hardcastle(2001)]{h01}Hardcastle, M.J., 2001, \aap, 373, 881
\bibitem[Hardcastle et al.(1998)Hardcastle, Birkinshaw \& Worrall]{hbw98}Hardcastle, M.J., Birkinshaw, M., \& Worrall, D.M., 1998, \mnras, 294, 615
\bibitem[Hardcastle et al.(2001a)Hardcastle, Birkinshaw \& Worrall]{hbw01a}Hardcastle, M.J., Birkinshaw, M., \& Worrall, D.M., 2001a, \mnras, 323, L17
\bibitem[Hardcastle et al.(2001b)Hardcastle, Birkinshaw \& Worrall]{hbw01b}Hardcastle, M.J., Birkinshaw, M., \& Worrall, D.M., 2001b, \mnras, 326, 1499
\bibitem[Hardcastle \& Worrall(1999)]{hw99}Hardcastle, M.J., \& Worrall, D.M., 1999, \mnras, 309, 969
\bibitem[Hardcastle \& Worrall(2000)]{hw00}Hardcastle, M.J., \& Worrall, D.M., 2000, \mnras, 319, 562
\bibitem[Harris et al.(1994)Harris, Carilli \& Perley]{hcp94}Harris, D.E., Carilli, C.L., \& Perley, R.A., 1994, \nat, 367, 713
\bibitem[Harris \& Krawczynski(2002)]{hk02}Harris, D.E., \& Krawczynski, H., 2002, \apj, 565, 244
\bibitem[Harris et al.(1998)Harris, Leighly \& Leahy]{hll98}Harris, D.E., Leighly, K.M., \& Leahy, J.P., 1998, \apj, 499, L149
\bibitem[Harris et al.(1999)]{hhss99}Harris, D.E., Hjorth, J., Sadun, A.C., Silverman, J.D., \& Vestergaard, M., 1999, \apj, 518, 213
\bibitem[Harris et al.(2000)]{hnpb00}Harris, D.E., et al., 2000, \apj, 530, L81
\bibitem[Harvanek et al.(2001)]{hesr01}Harvanek, M., Ellingson, E., Stocke, J.T., \& Rhee, G., 2001, \aj, 122, 2874
\bibitem[Heavens \& Meisenheimer(1987)]{hm87}Heavens, A.F., \& Meisenheimer, K., 1987, \mnras, 225, 335
\bibitem[Hill \& Lilly(1991)]{hl91}Hill, G.J., \& Lilly, S.J., 1991, \apj, 367, 1
\bibitem[Jester et al.(2002)]{jrmp02}Jester, S., R\"oser, H.-J., Meisenheimer, K., \& Perley, R., 2002, \aap, 385, L27
\bibitem[L\"ahteenm\"aki \& Valtaoja(1999)]{lv99}L\"ahteenm\"aki, A., \& Valtaoja, E., 1999, \aj, 117, 1168
\bibitem[Laing(1984)]{l84}Laing, R.A., 1984, in Bridle A.H., Eilek J.A., eds, Physics of Energy Transport in Radio Galaxies, NRAO Workshop no.~9, NRAO, Green Bank, West Virginia, p.~90
\bibitem[Laing et al.(1983)Laing, Riley \& Longair]{lrl83}Laing, R.A., Riley, J.M., \& Longair, M.S., 1983, \mnras, 204, 151
\bibitem[Leahy et al.(1989)Leahy, Muxlow \& Stephens]{lms89}Leahy, J.P., Muxlow, T.W.B., \& Stephens, P.W., 1989, \mnras, 239, 401
\bibitem[Leahy \& Perley(1991)]{lp91}Leahy, J.P., \& Perley, R.A., 1991, \aj, 102, 537
\bibitem[Leahy \& Perley(1995)]{lp95}Leahy, J.P., \& Perley, R.A., 1995, \mnras, 277, 1097
\bibitem[Lockman \& Savage(1995)]{ls95}Lockman, F.J., \& Savage, B.D., 1995, \apjs, 97, 1
\bibitem[Looney \& Hardcastle(2000)]{lh00}Looney, L.W., \& Hardcastle, M.J., 2000, \apj, 534, 172
\bibitem[Malaguti et al.(1994)Malaguti, Bassani \& Caroli]{mbc94}Malaguti, G., Bassani, L., \& Caroli, E., 1994, \apjs, 94, 517
\bibitem[Mathur et al.(1994)]{mwef94}Mathur, S., Wilkes, B., Elvis, M., \& Fiore, F., 1994, \apj, 434, 493
\bibitem[Meisenheimer et al.(1989)]{mrhy89}Meisenheimer, K., R\"oser, H.-J., Hiltner, P.R., Yates, M.G., Longair, M.S., Chini, R., \& Perley, R.A., 1989, \aap, 219, 63 
\bibitem[Meisenheimer et al.(1997)Meisenheimer, Yates \& R\"oser]{myr97}Meisenheimer, K., Yates, M.G., \& R\"oser, H.-J., 1997, \aap, 325, 57
\bibitem[Nicastro et al.(1999)]{nfpe99}Nicastro, F., Fiore, F., Perola, G.C., \& Elvis, M., 1999, \apj, 512, 136
\bibitem[Prieto(1997)]{p97}Prieto, M.A., 1997, \mnras, 284, 627
\bibitem[R\"oser(1989)]{r89}R\"oser, H.-J., 1989, in Meisenheimer K., R\"oser H.-J., eds, Hotspots in Extragalactic Radio Sources, Springer-Verlag, Heidelberg, p. 91
\bibitem[R\"oser \& Meisenheimer(1987)]{rm87}R\"oser, H.-J., \& Meisenheimer, K., 1987, \apj, 314, 70
\bibitem[Scheuer(1995)]{s95}Scheuer, P.A.G., 1995, \mnras, 277, 331
\bibitem[Schwartz et al.(2000)]{smlp00}Schwartz, D.A., et al., 2000, \apj, 540, L69
\bibitem[Stark et al.(1992)]{sgwb92}Stark, A.A., Gammie, C.F., Wilson, R.W., Bally, J., Linke, R.A., Heiles, C., \& Hurwitz, M., 1992, \apjs, 79, 77
\bibitem[Tashiro et al.(1998)]{tkmi98}Tashiro, M., et al., 1998, \apj, 499, 713
\bibitem[Tavecchio et al.(2000)]{tmsu00}Tavecchio, F., Maraschi, L., Sambruna, R.M., \& Urry, C.M., 2000, \apj, 544, L23
\bibitem[Tregillis et al.(2001)Tregillis, Jones \& Ryu]{tjr01}Tregillis, I.L., Jones, T.W., \& Ryu, D., 2001, \apj, 557, 475
\bibitem[Tribble(1991)]{t91}Tribble, P.C., 1991, \mnras, 253, 147
\bibitem[Tsakiris et al.(1996)]{tlsb96}Tsakiris, D., Leahy, J.P., Strom, R.G., \& Barber, C.R., 1996, in Ekers R.D., Fanti C., Padrielli L., eds, Extragalactic Radio Sources, IAU Symposium 175, Kluwer, Dordrecht, p.~256
\bibitem[Ueno et al.(1994)]{ukny94}Ueno, S., Koyama, K., Nishida, M., Yamauchi, S., \& Ward, M.J., 1994, \apj, 431, L1
\bibitem[Welch et al.(1996)]{wtpw96}Welch, W.J., et al., 1996, \pasp, 108, 93
\bibitem[Wilson et al.(2000)Wilson, Young \& Shopbell]{wys00}Wilson, A.S., Young, A.J., \& Shopbell, P.L., 2000, \apj, 544, L27
\bibitem[Wilson et al.(2001)Wilson, Young \& Shopbell]{wys01}Wilson, A.S., Young, A.J., \& Shopbell, P.L., 2001, \apj, 547, 740
\bibitem[Worrall et al.(2001a)Worrall, Birkinshaw \& Hardcastle]{wbh01a}Worrall, D.M., Birkinshaw, M., \& Hardcastle, M.J., 2001a, \mnras, 326, L7
\bibitem[Worrall et al.(2001b)]{wbhl01b}Worrall, D.M., Birkinshaw, M., Hardcastle, M.J., \& Lawrence, C.R., 2001b, \mnras, 326, 1127
\bibitem[Wu et al.(1999)Wu, Xue \& Fang]{wxf99}Wu, X.-P., Xue, Y.-P., \& Fang, L.-Z., 1999, \apj, 524, 22
\end{thebibliography}
\end{document}